\documentclass[pdflatex,sn-mathphys-num]{sn-jnl}

\usepackage{graphicx}%
\usepackage{multirow}%
\usepackage{amsmath,amssymb,amsfonts}%
\usepackage{amsthm}%
\usepackage{mathrsfs}%
\usepackage[title]{appendix}%
\usepackage{xcolor}%
\usepackage{textcomp}%
\usepackage{manyfoot}%
\usepackage{booktabs}%
\usepackage{algorithm}%
\usepackage{algorithmicx}%
\usepackage{algpseudocode}%
\usepackage{listings}%
\usepackage{longtable}
\usepackage{caption}

\DeclareMathOperator*{\argmin}{\arg\min}
\theoremstyle{thmstyleone}%
\newtheorem{thm}{Theorem}[section]
\newtheorem{lem}{Lemma}[section]
\newtheorem{assumption}{Assumption}[section]

\newcommand{\norm}[1]{\left\lVert#1\right\rVert}

\begin{document}
\title[High-dimensional Additive MAR]{High-Dimensional Regularized Additive Matrix Autoregressive Model}

\author[1]{\fnm{Debika} \sur{Ghosh}}\email{debika.2023phd@iimu.ac.in}

\author*[2]{\fnm{Samrat} \sur{Roy}}\email{samratr@iima.ac.in}

\author*[1]{\fnm{Nilanjana} \sur{Chakraborty}}\email{nilanjana.chakraborty@iimu.ac.in}

\affil[1]{\orgname{Indian Institute of Management Udaipur}}
\affil[2]{\orgname{Indian Institute of Management Ahmedabad}}


\abstract{High-dimensional time series has diverse applications in econometrics and finance. Recent models for capturing temporal dependence have employed a bilinear representation for matrix time series, or the Tucker-decomposition based representation in case of tensor time series. A bilinear or Tucker-decomposition based temporal effect is difficult to interpret on many occasions, along with its computational complexity due to the non-convex nature of the underlying optimization problem. Moreover, the existing matrix case models have not sufficiently explored the possibilities of imposing any lower-dimensional pattern on the transition matrices. In this work, we propose a regularized additive matrix autoregressive model with additive interaction of row-wise and column-wise temporal dependence, that offers more interpretability, less computational burden due to its convex nature and estimation of the underlying low rank plus sparse pattern of its transition matrices. We address the issue of identifiability of the various components in our model and subsequently develop a scalable Alternating Block Minimization algorithm for estimating the parameters. We provide a finite sample error bound under high-dimensional scaling for the model parameters. Finally, the efficacy of the proposed model is demonstrated on synthetic and real data. }


\maketitle

\section{Introduction}
\label{intro}
 High-dimensional time series models have gained a lot of prominence in recent years due to both technical developments (\cite{basu2015regularized}, \cite{zhang2017gaussian}, \cite{wang2022high}, \cite{adamek2023lasso}) and its various application areas, including finance and macroeconomics (\cite{de2008forecasting}, \cite{bernanke2005measuring}, \cite{blanchard2002empirical}), demography (\cite{gao2019high}), functional genomics (\cite{michailidis2013autoregressive}), dynamic traffic networks (\cite{chen2019modeling}) and neuroscience (\cite{seth2015granger}).
 
While most of the aforementioned work employed regularized versions of the Vector Autoregressive (VAR) model to capture underlying temporal dependence among vector-valued high-dimensional time series \citep{banbura2010large,basu2015regularized,kock2015oracle,ghosh2018high,wang2021high}, some recent studies have considered modeling temporal dependence among matrix-valued time series, wherein the observations at each time point are represented in the form of a matrix, and the interplays of its rows and columns are often sources of significant information. \cite{chen2021autoregressive} proposed such a matrix autoregressive (MAR) model in which they used a bilinear form $AY_{t-1}B^\prime$ to represent the temporal dependence between the data matrices $\{Y_t\}_{t=1}^T$, and the transition matrices $A$ and $B$ are aimed at capturing the row-wise and column-wise temporal dependence. Along the same line, \cite{li2021multi} considered a similar autoregressive model for tensor-variate time series (TAR), where they used a Tucker decomposed structure \citep{kolda2009tensor} to capture the underlying temporal dependence. To facilitate dimension reduction in the above-mentioned bilinear MAR model, both reduced rank structure \citep{xiao2022reduced} and sparsity structure \citep{hsu2021matrix} of the transition matrices have been explored. While these approaches help in reducing the dimensionality, they may suffer from the following problems:

\begin{enumerate}
    \item [(a)] In case of bilinear representation $AY_{t-1}B^\prime$, row-wise and column-wise temporal effects are convoluted in multiplicative interaction form, and it becomes difficult to disjoin and interpret the two effects separately \citep{zhang2024additive}. As illustrated in Section \ref{model}, while modeling the temporal dependence of matrix-valued macroeconomic data with different economic indicators across the rows and different countries across the columns, one may be interested in coherently estimating the two sources of temporal dependence $\textendash$ along different economic indicators, and along different countries; a bilinear convoluted structure will not serve that purpose.     
    \item [(b)] Though a reduced-rank or sparse structure imposed on the transition matrices $A$ and $B$ of the bilinear form $AY_{t-1}B^\prime$ alleviates the high-dimensionality of the parameters, it can be inadequate to represent the desired low-dimensional pattern on many occasions. For instance,  in the context of aforementioned macroeconomic matrix-variate data with economic indicators along the rows and countries along the columns, it is reasonable to assume that countries under the European Union follow harmonized economic and fiscal policies, and thus the temporal dependence pattern should be similar or `shared' across those countries. So, the transition matrix aimed at capturing the country-wise temporal effect, which is $B$ in this case, should ideally be a low-rank matrix. However, in case of bilinear form $AY_{t-1}B^\prime$, a low-rank B does not really characterize the aforementioned country-wise similar or `shared' temporal effect $\textendash$ for that, the representation $Y_{t-1} B^\prime$ would be more meaningful instead of the convoluted bilinear form $AY_{t-1}B^\prime$.            
    \item [(c)] Finally, with bilinear representation of the temporal dependence,  the estimation process becomes computationally involved $\textendash$ often the underlying optimization turns out to be a non-convex one.  
\end{enumerate}

In this paper, we propose a high-dimensional regularized additive matrix autoregressive model that overcomes the above-mentioned drawbacks. Our model captures the temporal dependence among the matrix-valued time series by employing an additive interaction form, wherein the overall temporal connection is represented as the sum of row-wise and column-wise temporal dependence in the data. To accommodate high-dimensionality of the parameters, we then impose different regularized structures on row-wise and column-wise transition matrices $\textendash$ low-rank, sparse, or low-rank plus sparse decomposed structure, depending on the context. As discussed in \cite{zhang2024additive}, this additive interaction form, as opposed to convoluted bilinear representation, offers more comprehensible interpretation of the row-wise and column-wise temporal dependence. Also, with additive form, the penalized transition matrices help in extracting meaningful low-dimensional pattern in the data, whereas, the same with bilinear form provides only dimension reduction. We develop a scalable alternating minimization algorithm to estimate the model parameters under high-dimensional setting that solves a convex optimization problem. We also address the issue of identifiability by employing a novel incoherence condition when low-rank plus sparse decomposed structure is imposed on the transition matrices. Finally, in terms of theoretical
 developments, we provide a detailed derivation and interpretation of the non-asymptotic upper bound of the estimation error under high dimensional scaling of the model parameters. To the best of our knowledge, the proposed methodology and the subsequent theoretical developments are novel contributions to the field of high-dimensional time series analysis.

 The remainder of the paper is organized as follows. Section \ref{model} provides a detailed description of our proposed model, illustrating all the steps involved in it, and also describes our algorithm to estimate the model parameters. Section \ref{theo} provides theoretical results related to the upper bound of the estimation error under high-dimensional scaling of the parameters. We then illustrate the performance of our posited method based on both synthetic and real data in Sections \ref{simu} and \ref{real_data} respectively, which is then followed by a concluding discussion in Section \ref{disc}. 


\section{Regularized Additive Matrix Autoregressive Model}
\label{model}
\begin{figure}[ht]
\includegraphics[scale=0.45]{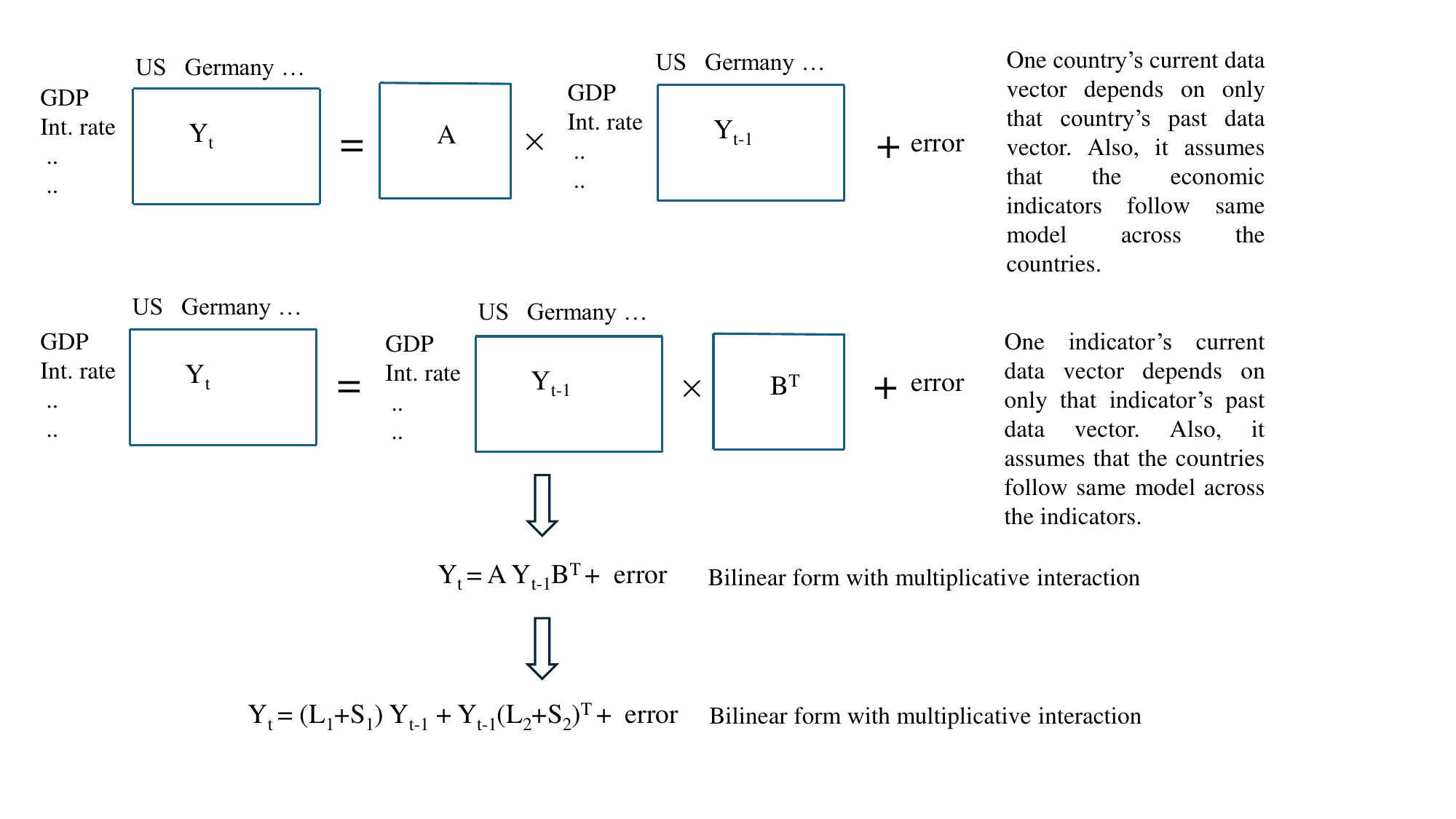}
\caption{Step-wise description of the proposed regularized additive matrix autoregressive model: First two models are the naive ones. The second one is the matrix autoregressive (MAR) model with multiplicative interaction of row-wise and column-wise temporal effect using a bilinear representation. The last one is our proposed additive matrix autoregressive model with additive interaction of row-wise and column-wise temporal dependence. Also, to deal with high-dimensionality, both row-wise and column-wise transition matrices are decomposed into a low rank and a sparse component.}
\label{fig:model_struc}
\end{figure}  
\subsection{Background}
\label{bground}
Suppose there are $d_1$ variables of interest, for $d_2$ entities, observed over T different time periods, and the objective is to model the underlying temporal dependence in the matrix valued time series $\{Y_t\in \mathbb{R}^{d_1 \times d_2}\}_{t=1}^T$. For example, the $d_1$ variables might represent different economic indicators -- such as Gross Domestic Product (GDP), Consumer Price Index (CPI), and others -- measured for $d_2$ different countries. As explained in \cite{chen2021autoregressive}, a naive approach to model such temporal dependence would be to employ a Vector Autoregressive (VAR) Model on the vectorized version of $Y_t$, which may fail to recognize the following intrinsic nature of $Y_t$ $\textendash$ there can be a strong temporal connection among the rows, that is, among the economic indicators (for any country), and similarly, there can be a strong temporal connection among the columns, that is, among the countries (for any economic indicator).

As depicted in Figure \ref{fig:model_struc}, an oversimplified model to capture the temporal dependence among $\{Y_t\}_{t=1}^T$ would be $Y_t= A\text{ }Y_{t-1}+E_t$ where $A\in \mathbb{R}^{d_1 \times d_1}$ is the transition matrix and $E_t$ is the error matrix at time point $t$. In this formulation, for any fixed country, $A$ captures the temporal connections among the economic indicators. However, this model suffers from the following drawbacks $\textendash$ first, each country's current data vector depends only on its own past data vector and thus the interactions among the countries (that is, among the columns) are not considered. Moreover, it assumes that the temporal dependence among economic indicators follows the same model across all the countries, which is indeed a restrictive assumption as the temporal dynamics of economic indicators of a developing country can differ significantly from those in a developed country. Similarly, another naive model would be $Y_t= Y_{t-1} B^\prime+E_t$ where, for any fixed economic indicator, $B^\prime\in \mathbb{R}^{d_2 \times d_2}$ reflects the temporal connections among the countries. However, in this case too, each indicator's current data vector depends only on its own past data vector and the interactions among the indicators (that is, among the rows) are not captured. Also, it is assumed that the temporal dependence among countries follows the same model across all the indicators $\textendash$ which is again a restrictive assumption as the underlying model capturing the temporal dynamics of GDP and CPI  may not be the same.

To overcome these limitations, \cite{chen2021autoregressive} combined the two above-mentioned oversimplified models and proposed a matrix autoregressive model where they used a bilinear form $AY_{t-1}B^\prime$ to capture the temporal dependence among the data matrices. In their framework, $A$ and $B$ capture the row-wise and column-wise temporal connections respectively, and interaction between rows and columns were modeled in a multiplicative form. When the number of parameters to estimate in $A$ and $B$ is higher than the number of observed data matrices $T$, one can impose regularized structure on $A$ and $B$ to deal with high-dimensionality. However, as mentioned earlier in Section \ref{intro}, a multiplicative interaction of row-wise and column-wise temporal dependence through the aforementioned bilinear form may face the following issues. Firstly, row-wise and column-wise temporal effects are convoluted in case of bilinear form, and as discussed in \cite{zhang2024additive}, it becomes difficult to disjoin and interpret the two effects separately. Furthermore, imposing low-dimensional patterns on $A$ and $B$ of the bilinear form is often insufficient to capture the underlying structure in the data. For instance, if the countries in the earlier example belong to the European Union, it is reasonable to assume that they follow harmonized economic and fiscal policies, and thus the temporal dependence pattern should be similar or `shared' across those countries. A natural approach to capture the above structure would be assuming low-rank structure on $B$. However, using a low-rank $B$ with $Y_{t-1} B^\prime$ to capture the above-mentioned pattern would be more meaningful rather than using a low-rank $B$ with $AY_{t-1} B^\prime$. Finally, in case of using a bilinear representation of the temporal dependence, the estimation process becomes computationally involved, often dealing with a non-convex optimization. 

\subsection{Regularized Additive MAR Model}
To address the aforementioned issues, we propose a regularized additive matrix autoregressive (MAR) model, where the primary step is to consider an additive interaction of row-wise and column-wise temporal dependence as follows:

\begin{equation}
    Y_t = AY_{t-1} + Y_{t-1} B^\prime + E_t, \text{ for } t=1,2,\cdots, T
    \label{main eqn 1}
\end{equation}
where $\{Y_t\in \mathbb{R}^{d_1 \times d_2}\}_{t=1}^T$ is a matrix-valued time series observed over $T$ time points, $A\in\mathbb{R}^{d_1 \times d_1}$ and $B\in\mathbb{R}^{d_2 \times d_2}$ are the transition matrices capturing row-wise and column-wise temporal dependence respectively, and $E_t \in \mathbb{R}^{d_1 \times d_2}$ is the error matrix at time point $t$. As discussed in \cite{chen2021autoregressive}, we assume that the error matrices $\{E_{t}\}_{t=1}^T$ are white noise in the sense that there is no correlation between $E_{t_1}$ and $E_{t_2}$ as long as $t_1\neq t_2$. However, $E_t$ is allowed to have any arbitrary correlations among its own elements. The simplest correlation structure one can consider on $E_t$ is to assume that the entries of $E_t$ are independent, implying that covariance matrix of $vec(E_t)$ is a diagonal matrix. On the other hand, as mentioned in \cite{chen2021autoregressive}, one can also consider a structured covariance matrix of $vec(E_t)$ as $\Sigma \in \mathbb{R}^{d_1d_2\times d_1d_2}$, where $\Sigma=\Sigma_1 \otimes I_{d_2}+I_{d_1}\otimes \Sigma_2$ and $\Sigma_1 \in \mathbb{R}^{d_1 \times d_1}$ and $\Sigma_2 \in \mathbb{R}^{d_2 \times d_2}$  are two symmetric positive semi-definite matrices.  

To alleviate the high-dimensionality of the model parameters in $A$ and $B$, one can assume different low-dimensional structures on them. Depending on the application at hand, one can assume $A$ and $B$ to be low-rank matrices, to be sparse matrices, or they can be assumed as decomposition of low-rank plus sparse matrices. To demonstrate the model and subsequent theoretical developments, we assume that $A$ and $B$ are decomposed as low-rank plus sparse matrices. In other words, $A=L_1+S_1$ and $B=L_2+S_2$, where $L_1\in \mathbb{R}^{d_1 \times d_1}$ and $L_2\in \mathbb{R}^{d_2 \times d_2}$ are the low-rank matrices and $S_1\in \mathbb{R}^{d_1 \times d_1}$ and $S_2\in \mathbb{R}^{d_2 \times d_2}$ are the sparse matrices, and the model in \eqref{main eqn 1} translates to
\begin{equation}
    Y_t = (L_1+S_1)Y_{t-1} + Y_{t-1} (L_2+S_2)^\prime + E_t, \text{ for } t=1,2,\cdots, T
    \label{main eqn 2}
\end{equation}
The assumption of low-rank plus sparse decomposed structure on the parameter matrices is quite common in the literature of high-dimensional data \citep{r1}. In our case, this implies that the underlying column-wise (and similarly, row-wise) temporal dependence will have two components $\textendash$ in the first component, the column-wise (and, row-wise) temporal dependence will be `similar' or `shared' across $d_2$ different entities (and, $d_1$ different variables). In addition to this baseline component of the column-wise (and row-wise) temporal dependence, there will be a second component where most of the column-wise (and row-wise) temporal effects will be zeros except for very few non-zero additional idiosyncratic temporal effects between the two entities (and between the two variables). For example, if the variables are different macroeconomic indicators and the entities are different countries in the European Union, then it is reasonable to assume that there will be a shared baseline component in the column-wise and row-wise temporal dependence as the countries in the European Union follow harmonized economic and fiscal policies in order to meet some common objectives and achieve an increased economic stability. On the other hand, the idiosyncratic components in the temporal dependence correspond to a financial crisis or an economic boom in some country, including Greece Government debt crisis, Portuguese financial crisis. Using the nuclear norm $\norm{\cdot}_{*}$ and $\ell_1$ norm $\norm{\cdot}_{1}$ (defined in Section 3) as suitable convex surrogates for low-rank and sparsity constraints respectively, now our aim is to minimize the following jointly convex objective function. 
{\footnotesize
\begin{equation}
\label{obj_func}
    \frac{1}{2T} \sum_{t=1}^{T} \left\lVert Y_t  - (L_1 + S_1)Y_{t-1} - Y_{t-1}(L_2+S_2)^\prime \right\rVert_{F}^{2} + \lambda_{S_1} \left\lVert S_1 \right\rVert_{1} + \lambda_{S_2} \left\lVert S_2 \right\rVert_{1} + \lambda_{L_1} \left\lVert L_1 \right\rVert_{*} + \lambda_{L_2} \left\lVert L_2 \right\rVert_{*}   
\end{equation}}

\noindent where $\lambda_{L_1}$, $\lambda_{L_2}$ and $\lambda_{S_1}$, $\lambda_{S_2}$ are non-negative regularization parameters for the low-rank and sparse components respectively. Later in Section \ref{theo}, we discuss the ideas to ensure identifiability of these low-rank and sparse components. 

\subsection{Estimation of the parameters}
Let us define the objective function in \eqref{obj_func} as $f(L_1,S_1,L_2,S_2)$. It is easy to verify that `$f$' is jointly convex in its arguments and hence the following alternating block  minimization procedure summarized in Algorithm \ref{Algo}, will obtain the desired minimizer.
\begin{algorithm}[H]
  \caption{Alternating Block Minimization for minimizing objective function: $f(L_1,S_1,L_2,S_2)$}
  \label{Algo}
  \begin{algorithmic}
    \State \textbf{Input}: data $\{Y_t\}_{t=1}^T, \lambda_{L_1}, \lambda_{L_2},\lambda_{S_1},\lambda_{S_2}$\label{op1}\\
    \State \textbf{Initialize}: $L_1^{(0)}, S_1^{(0)}, L_2^{(0)}, S_2^{(0)}$ \label{op2}\\
    \\
    \textbf{Repeat} 
    \State Step 1: Update $L_1^{(t+1)} = \underset{L_1}{\argmin} f(L_1, S_1^{(t)},L_2^{(t)}, S_2^{(t)})$ , given $S_1^{(t)},L_2^{(t)}, S_2^{(t)}$ \label{op3}\\
    \State Step 2: Update\text{ } $S_1^{(t+1)} = \underset{S_1}{\argmin} f(L_1^{(t+1)}, S_1,L_2^{(t)}, S_2^{(t)})$, given $L_1^{(t+1)}$, $L_2^{(t)}$, $S_2^{(t)}$\label{op4}\\
    \State Step 3: Update $L_2^{(t+1)} = \underset{L_2}{\argmin} f(L_1^{(t+1)}, S_1^{(t+1)},L_2, S_2^{(t)})$ , given $L_1^{(t+1)}, S_1^{(t+1)}, S_2^{(t)}$\label{op5}\\
    \State Step 4: Update \text{ }$S_2^{(t+1)} = \underset{S_2}{\argmin} f(L_1^{(t+1)}, S_1^{(t+1)},L_2^{(t+1)}, S_2)$ , given $L_1^{(t+1)}, S_1^{(t+1)}, L_2^{(t+1)}$\label{op6}\\
    \\
    \textbf{Until} {$f(L_1^{(t+1)}, S_1^{(t+1)},L_2^{(t+1)}, S_2^{(t+1)})$ converges.}
  \end{algorithmic}
\end{algorithm}  
In steps 1 and 3 of the above algorithm, we update the low-rank component $L_1$ and $L_2$ with nuclear norm penalization. This minimization problem shows up in various applications of machine learning, such as matrix classification, multi-task learning and matrix completion (see \cite{argyriou2008convex, tomioka2007classifying}). \cite{ji2009accelerated} considered a general class of optimization problems that includes the above formulation and proposed an Extended Gradient Algorithm and Accelerated Gradient Algorithm to obtain the minimizer. A direct application of the aforementioned algorithms provides the optimal solution in our case. On the other hand, in steps 2 and 4, when we update $S_1$ and $S_2$, we use the algorithm for penalized multivariate regression used in \cite{lin2016penalized}.

\section{Theoretical Results}
\label{theo}
We first define the estimation error $e^2(\hat{L}_1, \hat{L}_2, \hat{S}_1, \hat{S}_2)$ as given in \eqref{def}. In this section, we primarily focus on deriving a non-asymptotic upper bound to the estimation error. 

\begin{equation}
\label{def}
 e^2(\hat{L}_1, \hat{L}_2, \hat{S}_1, \hat{S}_2) = \left \lVert \hat{L}_{1} - L_1 \right \rVert_F^{2} + \left \lVert \hat{L}_{2} - L_2 \right \rVert_F^{2} +\left \lVert \hat{S}_{1} - S_1 \right \rVert_F^{2} + \left \lVert \hat{S}_{2} - S_2 \right \rVert_F^{2}.  
\end{equation}

\noindent
We first introduce some additional notations needed in the sequel.\\

\vspace{0.1in}
\noindent \textbf{Additional notation:} Let $R_1 \ll d_1$ and $R_2 \ll d_2$ denote the ranks of $L_1$ and $L_2$ respectively. We assume that $S_1$ and $S_2$ have $s_1 \ll d_1^2$ and $s_2 \ll d_2^2$ non-zero elements respectively. More specifically, suppose that $S_1$ is supported on a subset $E \subseteq \{1,2,\cdots,d_1^2\}$, with $|E|=s_1$. We define a pair of subspaces $(\mathbb{M}(E),\mathbb{M}^{\perp}(E))$, such that, $\mathbb{M}(E)=\{M\in \mathbb{R}^{d_1 \times d_1} \mid k^{th} \text{ element of } M =0, \forall k \notin E\}$ and $\mathbb{M}^{\perp}(E)= (\mathbb{M}(E))^{\perp}$. As shown in \cite{r1} and \cite{negahban2012unified}, one can easily verify that for any $M_1 \in \mathbb{M}(E)$ and $M_2 \in \mathbb{M}^{\perp}(E)$, $\norm{M_1+M_2}_{1}=\norm{M_1}_{1}+\norm{M_2}_{1}$. This ensures that the regularizer $\norm{\cdot}_{1}$ is decomposable (see \cite{negahban2012unified}) with respect to the subspace pair $(\mathbb{M}(E),\mathbb{M}^{\perp}(E))$. Simplifying the notation from $(\mathbb{M}(E),\mathbb{M}^{\perp}(E))$ to $(\mathbb{M},\mathbb{M}^{\perp})$, it is evident that, $S_1 \in \mathbb{M}$, $\pi_{\mathbb{M}} (S_1)= S_1$ and $\pi_{\mathbb{M}^{\perp}} (S_1)=0$, where $\pi_{\mathbb{M}}(\cdot)$ is the projection onto the subspace $\mathbb{M}$. We define $\hat{\Delta}_{L_1}=\hat{L}_1-L_1$, $\hat{\Delta}_{S_1}=\hat{S}_1-S_1$, $\hat{\Delta}_{L_2}=\hat{L}_2-L_2$ and $\hat{\Delta}_{S_2}=\hat{S}_2-S_2$.  Also, $\hat{\Delta}_{S_1}^{\mathbb{M}} = \pi_{\mathbb{M}}( \hat{\Delta}_{S_1})$ and $\hat{\Delta}_{S_1}^{\mathbb{M}^{\perp}} = \pi_{\mathbb{M}^{\perp}}( \hat{\Delta}_{S_1})$. Similarly, for a pair of subspaces $(\mathbb{N},\mathbb{N}^{\perp})$, we define $\hat{\Delta}_{S_2}^{\mathbb{N}} = \pi_{\mathbb{N}}( \hat{\Delta}_{S_2})$ and  $\hat{\Delta}_{S_2}^{\mathbb{N}^{\perp}} = \pi_{\mathbb{N}^{\perp}}( \hat{\Delta}_{S_2})$. The $\ell_1$ and $\ell_{\infty}$ norm of a matrix $A$ are defined by $\norm{A}_1=\underset{i}{\sum} \underset{j}{\sum}\lvert{a_{ij}}\rvert$ and $\norm{A}_{\infty}=\underset{i,j}{max}\lvert{a_{ij}}\rvert$ respectively. Denoting by $\sigma_1(A), \sigma_2(A),\cdots, \sigma_m(A)$, the singular values of $A\in \mathbb{R}^{m_1 \times m_2}$, where $m = min\{m_1, m_2\}$, we define the Nuclear Norm of $A$ by $\norm{A}_{*} = \sum_{j=1}^m \sigma_j(A)$ and the Spectral Norm of $A$ by $\norm{A}_{sp} = \underset{1\leq j \leq m}{max} \{\sigma_j(A)\}$. 
\\

\vspace{0.2in}
The roadmap for theoretical developments in this section is as follows: Lemmas \ref{lemma:31} and \ref{lemma:32} characterize the sets to which the errors ($\hat{\Delta}_{L_{1}}, \hat{\Delta}_{S_{1}})$ and $(\hat{\Delta}_{L_{2}},\hat{\Delta}_{S_{2}}$) belong. Later, on these sets, we assume Restricted Strong Convexity of the loss function (see Assumption \ref{ass 1}). For deterministic realizations of the errors, and under certain regularity conditions, Lemma \ref{lemma:33} establishes the bound on the estimation error $e^2(\hat{L}_1, \hat{L}_2, \hat{S}_1, \hat{S}_2)$. Theorem \ref{thm:34} extends the result to random realizations of the errors under Gaussian distribution.

\vspace{0.2in}

\begin{lem}
\label{lemma:31}
Let $C_1(L_1, S_1)$ and $C_2(L_2,S_2)$ be the weighted combinations of the nuclear norm and $\ell_1$ norm regularizers as follows:

\begin{align}
    & C_1(L_1, S_1)
    =  \left \lVert {L}_{1} \right \rVert _{*} + \frac{\lambda_{{S_1}}}{\lambda_{{L_1}}} \left \lVert {S}_{1} \right \rVert _{1} \nonumber \\
    & C_2(L_2, S_2)
    =  \left \lVert {L}_{2} \right \rVert _{*} + \frac{\lambda_{{S_2}}}{\lambda_{{L_2}}} \left \lVert {S}_{2} \right \rVert _{1}
    \label{defn C1 C2}
\end{align}

Then, for any $R_1 = 1,2\dots d_1$ and $R_2 = 1,2 \dots d_2$, there exists decomposition of the forms $\hat{\Delta}_{L_{1}} = \hat{\Delta}_{
L_{1}}^{A_1} + \hat{\Delta}_{L_{1}}^{B_1} $ and
$\hat{\Delta}_{L_{2}} = \hat{\Delta}_{
L_{2}}^{A_2} + \hat{\Delta}_{L_{2}}^{B_2} $
with
\textit{rank}$\big(\hat{\Delta}_{L_{1}}^{A_1}$\big) $\leq 2R_1$,
\textit{rank}$\big(\hat{\Delta}_{L_{2}}^{A_2}$\big) $\leq 2R_2$, $L_1^T\hat{\Delta}_{L_{1}}^{B_1} =0 $, $L_1(\hat{\Delta}_{L_{1}}^{B_1})^T =0$, $L_2^T \hat{\Delta}_{L_{2}}^{B_2} =0 $, $L_2(\hat{\Delta}_{L_{2}}^{B_2})^T =0$
and

\begin{flalign}
    & C_1(L_1, S_1) - C_1(L_1 + \hat{\Delta}_{L_{1}}, 
    S_1 + \hat{\Delta}_{S_{1}})
    \leq C_1(\hat{\Delta}_{L_{1}}^{A_1}, \hat{\Delta}_{S_{1}}^\mathbb{M}) - C_1(\hat{\Delta}_{L_{1}}^{B_1}, \hat{\Delta}_{S_{1}}^\mathbb{M^{\bot}})
    \label{eq2}
\end{flalign}
\begin{flalign}
    & C_2(L_2,S_2) - C_2(L_2 + \hat{\Delta}_{L_{2}}, 
    S_2 + \hat{\Delta}_{S_{2}})
    \leq
    C_2(\hat{\Delta}_{L_{2}}^{A_2}, \hat{\Delta}_{S_{2}}^\mathbb{N}) - C_2(\hat{\Delta}_{L_{2}}^{B_2}, \hat{\Delta}_{S_{2}}^\mathbb{N^{\bot}})
    \label{eq3}
\end{flalign}
\end{lem} 
\vspace{0.2in}
\begin{lem}
\label{lemma:32}
Suppose that the errors $E_t$ are deterministic. Let $\mathcal{D}_1$ and $\mathcal{D}_2$ be the matrices defined as follows:
\begin{align}
    & \mathcal{D}_1 = \frac{1}{T} \sum_{t=1}^{T} E_t Y_{t-1}^T \nonumber \\  
     & \mathcal{D}_2 = \frac{1}{T} \sum_{t=1}^{T} E_t^{T} Y_{t-1} \nonumber 
\end{align}
Then, under the conditions $\lambda_{L_1} \geq 4 \left \lVert \mathcal{D}_1 \right\rVert_{sp}$, $\lambda_{L_2} \geq 4 \left \lVert \mathcal{D}_2 \right\rVert_{sp}$, $\lambda_{S_1} \geq 4 \left \lVert \mathcal{D}_1 \right\rVert_{\infty}$ and $\lambda_{S_2} \geq 4 \left \lVert \mathcal{D}_2 \right\rVert_{\infty}$, the errors ($\hat{\Delta}_{L_{1}}, \hat{\Delta}_{S_{1}})$ and $(\hat{\Delta}_{L_{2}},\hat{\Delta}_{S_{2}}$) will satisfy the following constraints:
\begin{align}
\label{rsset}
    C_1(\hat{\Delta}_{L_{1}}^{B_1}, \hat{\Delta}_{S_{1}}^\mathbb{M^{\bot}}) \leq 3 C_1(\hat{\Delta}_{L_{1}}^{A_1}, \hat{\Delta}_{S_{1}}^\mathbb{M}) \nonumber \\
    C_2(\hat{\Delta}_{L_{2}}^{B_2}, \hat{\Delta}_{S_{2}}^\mathbb{N^{\bot}}) \leq 3 C_2(\hat{\Delta}_{L_{2}}^{A_2}, \hat{\Delta}_{S_{2}}^\mathbb{N})
\end{align}    
\end{lem}

As mentioned earlier, the above lemmas characterize the sets in which the errors ($\hat{\Delta}_{L_{1}}, \hat{\Delta}_{S_{1}})$ and $(\hat{\Delta}_{L_{2}},\hat{\Delta}_{S_{2}}$) lie. Given this set, we are now in a position to summarize all the assumptions that we make. We first prepare a list of the assumptions and then provide further details on each of those assumptions.


\begin{assumption}
 \label{ass 1} 
The loss function $\frac{1}{2T} \sum_{t=1}^{T} \left\lVert Y_t - (L_1 + S_1)Y_{t-1} - Y_{t-1}(L_2+S_2)^T \right\rVert_{F}^{2}$, denoted by $L(L_1, L_2, S_1, S_2)$, satisfies the \textit{Restricted Strong Convexity} condition with curvature $\gamma > 0$ over the set characterized by Lemma \ref{lemma:31} and Lemma \ref{lemma:32}. In other words, there exists a positive constant $\gamma>0$ such that 

\begin{flalign}
\label{rsceq}
    &\frac{1}{2T} \sum_{t=1}^{T} \left\lVert \big[ \hat{\Delta}_{L_{1}} + \hat{\Delta}_{S_{1}} \big] Y_{t-1} + Y_{t-1} \big[ \hat{\Delta}_{L_{2}} + \hat{\Delta}_{S_{2}} \big]^{T} \right\rVert _{F}^{2} \nonumber \\
    & \geq \frac{\gamma}{2} \Bigg[ \left \lVert \hat{\Delta}_{L_{1}}+ \hat{\Delta}_{{S_{1}}}  \right \rVert ^{2}_{F} + \left \lVert \hat{\Delta}_{{L_{2}}} + \hat{\Delta}_{{S_{2}}}  \right \rVert ^{2}_{F}\Bigg]
\end{flalign}   
\end{assumption}
\vspace{0.2in}
\begin{assumption}
\label{ass 2}    
\begin{equation}
    \left\lVert L_1 \right\rVert_{\infty} \leq \frac{\alpha_1}{\sqrt{d_1 d_1}} \text{ and } \left\lVert L_2 \right\rVert_{\infty} \leq \frac{\alpha_2}{\sqrt{d_2 d_2}}.    
\end{equation}
for some fixed parameter $\alpha_1$ and $\alpha_2$.
\end{assumption}
\vspace{0.2in}
\begin{assumption}
\label{ass 3}
 When the errors $E_{t}$ are deterministic, the regularization parameters $(\lambda_{L_1}, \lambda_{L_2}, \lambda_{S_1}, \lambda_{S_2} )$ satisfy the following constraints:

\begin{align}
    & \lambda_{L_1} \geq 4 \left \lVert \mathcal{D}_1 \right\rVert_{sp}, \hspace{0.5cm}
    \lambda_{S_1} \geq 4 \left \lVert \mathcal{D}_1 \right\rVert_{\infty} + \frac{4\gamma \alpha_1}{\sqrt{d_1 d_1}} \nonumber \\ \label{eq4}  
    & \lambda_{L_2} \geq 4 \left \lVert \mathcal{D}_2 \right\rVert_{sp}, \hspace{0.5cm}
    \lambda_{{S_2}} \geq 4 \left \lVert \mathcal{D}_2 \right\rVert_{\infty} + \frac{4\gamma \alpha_2}{\sqrt{d_2 d_2}} 
\end{align}
where $\mathcal{D}_1$ and $\mathcal{D}_2$ are the same as defined in Lemma \ref{lemma:32}.
\end{assumption}

\begin{itemize}
\item Assumption \ref{ass 1} ensures that the loss function exhibits strong convexity over some restricted set of interest. In other words, this implies that the loss function should have sharp curvature around the optimal solution, ensuring that a small difference in loss implies a small error. Otherwise, if there is not sufficient curvature of the loss function around the optimal solution, then the error can be large even if the difference in loss is small. The latter is undesirable and that is ameliorated by imposing the condition given in \eqref{rsceq} (see \cite{negahban2012unified}). Note that, it is impossible to ensure global strong convexity under high-dimensional setup and thus, a common practice is to ensure strong convexity on some `restricted set' of interest \citep{negahban2012unified}. In our case, as derived earlier in Lemma \ref{lemma:31} and Lemma \ref{lemma:32}, that set is essentially characterized by \eqref{rsset}, and thus Assumption \ref{ass 1} ensures strong convexity of the loss function for that restricted set. This is a fairly standard assumption in the high-dimensional literature \cite{r1, roy2022regularized}.

\item Assumption \ref{ass 2} is aimed to ensure that the low-rank components $L_1$ and $L_2$ are incoherent with the sparse components $S_1$ and $S_2$ respectively. This assumption is a straightforward application of the `spikiness' restriction on the the low-rank matrix, as introduced in \cite{r1}. As described in \cite{r1}, when the parameter $\alpha_1$ (and similarly $\alpha_2$) $\approx$ 1, then all the `mass' of $L_1$ (and of $L_2$) is distributed equally among its $d_1^2$ (or, $d_2^2$) elements, which is a case of `minimal spikiness' of $L_1$ (and $L_2$). On the other extreme, when the parameter $\alpha_1 \approx \sqrt{d_1d_1}$ (or, $\alpha_2 \approx \sqrt{d_2d_2}$), then all the mass of $L_1$ (or, of $L_2$) will be concentrated only on one element and the other elements will be zeros. In this latter case, $L_1$ and $L_2$ will have `maximal spikiness', implying that they will essentially become sparse matrices, which is undesirable. Thus, by controlling the spikiness of the low-rank matrices, the parameters $\alpha_1$ and $\alpha_2$ ensure sufficient incoherence between the low-rank and sparse components. In practice, the values of $\alpha_1$ and $\alpha_2$ are set between the above two extremes. This incoherence assumption is milder than other incoherence conditions in the existing literature, including those in \cite{liu2019low,zhang2016exact}, which involve the components of SVD.                            
\item Assumption \ref{ass 3} imposes certain lower bounds to the regularizer parameters, which is a common requirement in the high-dimensional literature.
\end{itemize}

\vspace{0.2in}
Under the above assumptions, the following lemma establishes an upper bound to  $e^2(\hat{L}_1, \hat{L}_2, \hat{S}_1, \hat{S}_2)$ in the case of deterministic errors.
\begin{lem}
\label{lemma:33}
Suppose the errors $E_t$ are deterministic. Then, under Assumptions \ref{ass 1}, \ref{ass 2}, \ref{ass 3}, the estimation error satisfies the following condition:
\begin{align}
    &  e^2(\hat{L}_1, \hat{L}_2, \hat{S}_1, \hat{S}_2)\preceq  \lambda_{{L}_1}^2 R_1 + \lambda_{{S}_1}^2 s_1 + \lambda_{{L}_2}^2 R_2+  \lambda_{{S}_2}^2 s_2
    \end{align}
where the notation '$\preceq$' denotes an upper bound, ignoring all constant factors. \\    
\end{lem}

Note that the above result is broadly in line with Theorem 1 in \cite{r1}; specifically, when the loss function satisfies the Restricted Strong Convexity and the parameters of interest are exactly (not approximately) low-rank and sparse, a similar form of error bound is obtained in \cite{r1}. In our setting, we have two low-rank and two sparse components and thus, there are four terms corresponding to each component in the above bound.

\vspace{0.2in}
We now extend the above result under a Gaussian distribution assumption on the errors. To that end, define $\mathbb{E}_1$ as a data matrix of order $d_1 \times T d_{2}$, constructed by arranging the time series $\{E_t\}_{t=1}^T$ side by side. Similarly, let $Y_{-1}^{(1)}$ be a data matrix of order $d_1 \times T d_{2}$, formed by arranging the time series $\{Y_{t-1}\}_{t=1}^T$ side by side. Next, define $\mathbb{E}_2$ as a data matrix of order $Td_1 \times d_{2}$, created by stacking the time series $\{E_t\}_{t=1}^T$ one below the other. Similarly, let $Y_{-1}^{(2)}$ be a data matrix of order $T d_1 \times d_{2}$, formed by stacking the time series $\{Y_{t-1}\}_{t=1}^T$ one below the other. It is easy to verify that the matrices $\mathcal{D}_1$ and $\mathcal{D}_2$, defined earlier in Lemma \ref{lemma:32}, can be expressed as 
$\mathcal{D}_1 = \frac{1}{T} \mathbb{E}_1 Y_{-1}^{(1)^{T}}$ and $\mathcal{D}_2 = \frac{1}{T} \mathbb{E}_2^{T} Y_{-1}^{(2)}$.

Now, let $\{p_{1t}\}$ be a process characterized by the columns of $\mathbb{E}_1$, which is a centered, stationary, Gaussian process. Similarly, let $\{p_{2t}\}$ be a process characterized by the columns of $Y_{-1}^{(1)}$. It is assumed that, the process $\{p_{2t}\}$ is also a centered, stationary, Gaussian process, and it is obvious that $Cov(p_{1t},p_{2t})=0\: \forall t$. As in \cite{basu2015regularized}, we first define the spectral density corresponding to the process $\{p_{1t}\}$ as follows $f_{p_1}(\theta)= \frac{1}{2\pi} \sum_{\ell=-\infty}^{\infty}\Gamma_{p_1}(\ell) e^{-i \ell\theta}, \theta\in[-\pi,\pi]$, and assume that it exists with its maximum eigen value being bounded a.e. on $[-\pi,\pi]$. In terms of notation, this implies that $\mathscr{M}(f_{p_1})=\underset{\theta \in [-\pi,\pi]}{\text{ess sup}} \Lambda_{\text{max}}(f_{p_1}(\theta))$ $< \infty$. Similarly, the maximum eigen value of the spectral density corresponding to the process $\{p_{2t}\}$ is denoted by $\mathscr{M}(f_{p_2})$ and we assume that $\mathscr{M}(f_{p_2}) < \infty$. Finally, we define the cross spectral density of the two processes $\{p_{1t}\}$ and $\{p_{2t}\}$ as $f_{p_1,p_2}(\theta)= \frac{1}{2\pi} \sum_{\ell=-\infty}^{\infty}\Gamma_{p_1,p_2}(\ell) e^{-i \ell\theta}, \theta\in[-\pi,\pi]$ where $\Gamma_{p_1,p_2}(h)=Cov(p_{1t},p_{2\text{ }\overline{t+h}}), \text{ }t,h \in \mathbb{Z}$. We assume that the above cross spectral density exists and its maximum eigen value is bounded a.e. on $[-\pi,\pi]$. In terms of the notation, $\mathscr{M}(f_{p_1,p_2})=\underset{\theta \in [-\pi,\pi]}{\text{ess sup }} \sqrt{\Lambda_{\text{max}}(f^{*}_{p_1,p_2}(\theta) f_{p_1,p_2}(\theta))}$ $< \infty$. We then define $Q_1$ as 

\begin{equation}
    Q_1 =  \mathscr{M}(f_{p_1})+ \mathscr{M}(f_{p_2})+\mathscr{M}(f_{p_1,p_2}).
\end{equation}

Similarly, let $\{q_{1t}\}$ be a process characterized by the rows of $\mathbb{E}_2$, which is a centered, stationary, Gaussian process. Also, let $\{q_{2t}\}$ be a process characterized by the rows of $Y_{-1}^{(2)}$. It is assumed that, the process $\{q_{2t}\}$ is also a centered, stationary, Gaussian process, and it is obvious that $Cov(q_{1t},q_{2t})=0\: \forall t$. As before, we first define the spectral density corresponding to the process $\{q_{1t}\}$ as follows $f_{q_1}(\theta)= \frac{1}{2\pi} \sum_{\ell=-\infty}^{\infty}\Gamma_{q_1}(\ell) e^{-i \ell\theta}, \theta\in[-\pi,\pi]$, and assume that it exists with its maximum eigen value being bounded a.e. on $[-\pi,\pi]$. In terms of notation, this implies that $\mathscr{M}(f_{q_1})=\underset{\theta \in [-\pi,\pi]}{\text{ess sup}} \Lambda_{\text{max}}(f_{q_1}(\theta))$ $< \infty$. Similarly, the maximum eigen value of the spectral density corresponding to the process $\{q_{2t}\}$ is denoted by $\mathscr{M}(f_{q_2})$ and we assume that $\mathscr{M}(f_{q_2}) < \infty$. Finally, we define the cross spectral density of the two processes $\{q_{1t}\}$ and $\{q_{2t}\}$ as $f_{q_1,q_2}(\theta)= \frac{1}{2\pi} \sum_{\ell=-\infty}^{\infty}\Gamma_{q_1,q_2}(\ell) e^{-i \ell\theta}, \theta\in[-\pi,\pi]$ where $\Gamma_{q_1,q_2}(h)=Cov(q_{1t},q_{2\text{ }\overline{t+h}}), \text{ }t,h \in \mathbb{Z}$. We assume that the above cross spectral density exists and its maximum eigen value is bounded a.e. on $[-\pi,\pi]$. In terms of the notation, $\mathscr{M}(f_{q_1,q_2})=\underset{\theta \in [-\pi,\pi]}{\text{ess sup }} \sqrt{\Lambda_{\text{max}}(f^{*}_{q_1,q_2}(\theta) f_{q_1,q_2}(\theta))}$ $< \infty$. We then define $Q_2$ as 

\begin{equation}
    Q_2 =  \mathscr{M}(f_{q_1})+ \mathscr{M}(f_{q_2})+\mathscr{M}(f_{q_1,q_2}).
\end{equation}
\vspace{0.1in}
\begin{thm}
\label{thm:34}
Suppose that $vec(E_t)$ are i.i.d. MVN$(0, \Sigma)$, where $\Sigma = \big[ \Sigma_1 \otimes I_{d_2} + I_{d_1} \otimes \Sigma_2  \big]$, and also assume that Assumption \ref{ass 2} holds. Then it can be shown that conditions in Assumptions \ref{ass 1} and \ref{ass 3} are satisfied with high probability and we will have 
\begin{eqnarray*}
 e^2(\hat{L}_1, \hat{L}_2, \hat{S}_1, \hat{S}_2) &\leq& s_1\{c_1Q_1^2\frac{2\log d_1}{T}+c_2\frac{\gamma^2\alpha_1^2}{d_1^2}\}+ s_2\{c_3Q_2^2\frac{2\log d_2}{T}+c_4\frac{\gamma^2\alpha_2^2}{d_2^2}\}+\\
 &&c_5 Q_1^2 R_1 \frac{2d_1}{T} + c_6 Q_2^2 R_2 \frac{2d_2}{T}.  
\end{eqnarray*}
with probability $1- max(e^{-c_1 log(d_1)}, e^{-c_2 log(d_2)})$ for some suitably chosen constants $c_1, c_2, c_3, c_4, c_5$ and $c_6$.
\end{thm}
\vspace{0.1in}
The above bound is analogous to the ones obtained in the existing literature. The terms $s_1Q_1^2\frac{2\log d_1}{T}$ and $s_2Q_2^2\frac{2\log d_2}{T}$ are in line with the sparse regularized vector autoregressive case \cite{basu2015regularized}. These terms can be interpreted as follows: the term $s_1Q_1^2\frac{2\log d_1}{T}$ arises as a result of estimating $s_1$ non-zero elements of $d_1 \times d_1$ dimensional matrix $S_1$. Note that, there are ${d_1^2 \choose s_1}$ possible subsets of size $s_1$ and thus the numerator includes the corresponding term with the scaling $\log({d_1^2 \choose s_1})\approx s_1 2\log(d_1)$. A similar interpretation follows for the term $s_2Q_2^2\frac{2\log d_2}{T}$. The terms $Q_1^2R_1\frac{2d_1}{T}$ and $Q_2^2R_2\frac{2d_2}{T}$ are in line with \cite{r1} and \cite{roy2022regularized}, where $R_1\times 2d_1$ and $R_2\times 2d_2$ correspond to the number of free elements in $L_1$ and $L_2$ respectively. Finally, the terms $\frac{s_1\gamma^2\alpha_1^2}{d_1^2}$ and $\frac{s_2\gamma^2\alpha_2^2}{d_2^2}$ appear because of non-identifiability of the low-rank and sparse components.     

\section{Performance Evaluation}
\label{simu}
In this section, we evaluate the performance of our proposed method based on synthetic data under different settings. As mentioned in Section \ref{model}, one can assume various low-dimensional structure for row-wise and column-wise transition matrices of our additive MAR model $\textendash$ low-rank, sparse, low-rank plus sparse. For ease of exposition, we first assess the estimation quality of our model separately with low-rank transition matrices, and with sparse transition matrices in Section \ref{Sim:est}. Additionally, in Section \ref{Sim:pred} we evaluate the model’s predictive performance assuming a low-rank plus sparse decomposition of the transition matrices. This holistic approach allows us to explore the performance of our model under different regularization of the transition matrices.          
\vspace{0.07in}\\

\noindent
\subsection{Estimation quality} \label{Sim:est}
\noindent\textit{Data generating process:} We begin by describing the procedure for generating the true low-rank components ${L_1}$,  ${L_2}$ and the true sparse components ${S_1}$, ${S_2}$ of our model. To generate ${L_1} \in \mathbb{R}^{d_1 \times d_1}$ with rank $R_1$, we first start with a matrix in $\mathbb{R}^{d_1 \times d_1}$ with entries from Uniform (0,1), and then obtain its singular value decomposition (SVD). We then randomly select $(d_1-R_1)$ diagonal elements of the diagonal matrix $D$ of the above-mentioned SVD, change those elements to zeros while the others remain non-zeros, and name the resulting matrix as ${D_1}$. Finally, the matrix ${L_1}$ with rank $R_1$ can be generated as ${U} {D_1} {V}^T$, where ${U}$ and ${V}$ are the matrices with orthonormal columns from the aforementioned SVD. The matrix ${L_2} \in \mathbb{R}^{d_2 \times d_2}$ with rank $R_2$ can be generated in a similar fashion.    

To generate the sparse components, we first start with a matrix with all its elements as zeros, then randomly select a small proportion of the elements and replace those zeros with entries from Uniform distribution, whose range is governed by a pre-specified maximum eigenvalue that controls the spectral properties of the matrix. Then the signs of those non-zero elements are decided by tossing a fair coin. The above-mentioned proportion of non-zero elements in the sparse components is referred to as edge-density. Finally, to ensure the stationarity of the generated matrix, we check its maximum absolute eigen value, and if the same is higher than the above-mentioned pre-specified value, we scale down the entries of the matrix in such a way that the condition is satisfied.

Given the true low-rank and sparse transition matrices, we generate the error matrices ${\{E_t\in\mathbb{R}^{d_1 \times d_2}\}_{t=1}^T}$, where, as mentioned earlier in Sections \ref{model} and \ref{theo}, $vec({E_t})$ are drawn independently and identically from a Multivariate Normal distribution with mean zero and covariance matrix ${\Sigma}$, where ${\Sigma} = \big[ {\Sigma_1} \otimes {I}_{d_2} + {I}_{d_1} \otimes {\Sigma_2}  \big]$ and ${\Sigma_1\in \mathbb{R}^{d_1 \times d_1}}$ and ${\Sigma_2\in \mathbb{R}^{d_2 \times d_2}}$ are two symmetric positive semi-definite matrices. Finally, the data matrices ${Y_t \in \mathbb{R}^{d_1 \times d_2}}$ are generated recursively $\textendash$ that is, ${Y_t=L_1 Y_{t-1}+Y_{t-1}L_2^\prime+E_t}$ when the transitions matrices are assumed to have low-rank structure, or, ${Y_t=S_1 Y_{t-1}+Y_{t-1}S_2^\prime+E_t}$ when the transitions matrices are assumed to have sparse structure. We then employ our proposed algorithm in Section \ref{model} on this simulated data to obtain the estimates. The regularization parameters $\lambda_{{L_1}}$, $\lambda_{{L_2}}$, $\lambda_{{S_1}}$ and $\lambda_{{S_2}}$ are selected using a grid search method. More specifically, in case of low-rank transition matrices, we run the algorithm and obtain estimates of ${L_1}$ and ${L_2}$ for different grids of the pair ($\lambda_{{L_1}}$, $\lambda_{{L_2}}$) and select that pair for which the ranks of the estimated low-rank components are as close as possible to the ranks of the true ${L_1}$ and ${L_2}$, that is ${R_1}$ and ${R_2}$ respectively. Similarly, for the sparse transition matrices, the optimal choices for ($\lambda_{{S_1}}$, $\lambda_{{S_2}}$) are those for which the numbers and positions of the zero and non-zero elements in the estimated sparse components are as close as possible to the same in the true sparse components. Later in this section, we develop an AIC criteria, which facilitates selection of the optimum values of the regularization parameters when the true ranks and sparsity levels are unknown to us.\\

\noindent \textit{Evaluation criteria}: We primarily use the notion of Relative Error to evaluate the estimation quality of our proposed method. In case of low-rank structure of the transition matrices, Relative Error (RE) is defined as 
$$\frac{\left \lVert {\hat{L}_{1}} - {L_1} \right \rVert_F^{2} + \left \lVert {\hat{L}_{2}} - {L_2} \right \rVert_F^{2}} { \left \lVert{L_1} \right\rVert_F^{2} +  \left \lVert{L_2} \right\rVert_F^{2}}.$$
Similarly, for the sparse structure of the transition matrices, Relative Error (RE) is defined as 
$$\frac{\left \lVert {\hat{S}_{1}} - {S_1} \right \rVert_F^{2} + \left \lVert {\hat{S}_{2}} - {S_2} \right \rVert_F^{2}} { \left \lVert{S_1} \right\rVert_F^{2} +  \left \lVert{S_2} \right\rVert_F^{2}}.$$
The quality of the estimation is indicated by low values of the above relative errors and the similarity in rank between estimated transition matrices ${\hat{L}_{1}}$, ${\hat{L}_{2}}$ and the true parameters ${L_1}$, ${L_2}$. Alongside that, the measures of sensitivity and specificity help to assess the effectiveness of support recovery for the estimation of the sparse components $S_1$ and $S_2$, which are defined as follows
\begin{enumerate}
    \item Specificity for $\hat{S_1}$, denoted by $SP_{S_1}$, is defined as the proportion of true negatives or alternatively 1 - False Positive Rate (FPR), where, FPR is defined as
    $$
    \frac{\text{Total number of non-zero elements in} \hspace{0.1 cm} {\hat{S}_1} \hspace{0.1 cm} \text{that are actually zero in} \hspace{0.1 cm} {S_1}   }{\text{Total number of elements in} \hspace{0.1 cm} {S_1} \hspace{0.1 cm} \text{that are actually zero}}
    $$
    \item Sensitivity for $\hat{S_1}$, denoted by $SN_{S_1}$, is defined as the True Positive Rate (TPR) as follows 
    $$
    \frac{ \text{Total number of non-zero elements in} \hspace{0.1cm} {S_1} \hspace{0.1cm} \text{that are correctly classified as non-zero in} \hspace{0.1cm} {\hat{S_1}}  }{\text{Total number of elements in} \hspace{0.1 cm} {S_1} \hspace{0.1 cm} \text{that are actually non-zero}}
    $$
\end{enumerate}
$SP_{S_2}$ and $SN_{S_2}$ are defined in a similar way. Higher values of SP and SN, that is values either close to 1 or exactly 1, are preferable.\\

\noindent \textit{Numerical Results}: We now assess the performance of our model using the above-mentioned metrics under different setup. Each setup here corresponds to a specific combination of the pair ($d_1$, $d_2$). Additionally, under each setup we have different sub-cases denoting the varying levels of sparsity and different true rank values for the model with sparse regularization and low rank regularization respectively. \\
Different setups for the model with sparse regularization include the following: 

\begin{itemize}
    \item Setup 1: $d_1$ = 15, $d_2$ = 10; Setup 2: $d_1$ = 30, $d_2$ = 20  
    \item Sub-case 1:  $e_1$ = 0.2, $e_2$ = 0.2; Sub-case 2:  $e_1$ = 0.4, $e_2$ = 0.4, where $e_1$ and $e_2$ are the edge densities of $S_1$ and $S_2$ respectively. 
\end{itemize}
\noindent

Likewise, different setups for the model with low rank regularization are as follows:

\begin{itemize}
    \item  Setup 3: $d_1$ = 15, $d_2$ = 10; Setup 4: $d_1$ = 30, $d_2$ = 20
    \item Sub-case 1:  $R_1$ = 3, $R_2$ = 3; Sub-case 2:  $R_1$ = 5, $R_2$ = 5, where $R_1$ and $R_2$, as defined earlier, are the ranks of $L_1$ and $L_2$ respectively.
\end{itemize}
Performance evaluation results obtained for each of the two models, that is models with low-rank regularization and sparsity regularization, are summarized in the following tables under the aforementioned setups. It is evident from Table \ref{tab: eg1} and Table \ref{tab: eg2} that as the number of time points increases, the relative error decreases. Alongside that, we also see that better support recovery, that is higher sensitivity and specificity, is achieved with higher values of $T$. It is obvious that, both relative errors and support recovery measures are in general slightly better in Table \ref{tab: eg1} as compared to Table \ref{tab: eg2} as the setup in Table \ref{tab: eg2} has higher burden in terms of parameters. Thus, slightly higher values of $T$ would make the estimation quality in Table \ref{tab: eg2} as good as in Table \ref{tab: eg1}. Similar pattern is also observed in Table \ref{tab:eg3} and Table \ref{tab:eg4}. Finally, it is worth noting that for any fixed setup in Table \ref{tab: eg1} and Table \ref{tab: eg2}, when edge density is increased from 0.2 to 0.4, there is an increase in the relative error. Similarly, for any fixed setup in Table \ref{tab:eg3} and Table \ref{tab:eg4}, when true ranks $R_1$ and $R_2$ are increased, relative error also increases. This finding is consistent with the expression of the estimation error bound obtained in Theorem \ref{thm:34}.
\begin{table}[h]
\caption{Performance Evaluation under setup 1: $d_1 = 15$, $d_2 = 10$. Relative error, sensitivity and specificity are reported for two different sparsity levels, that is, edge densities 0.2 and 0.4. As the number of time points increases, estimation quality improves. Also, for any fixed time point, when the edge density increases from 0.2 to 0.4, the relative error increases, which is in line with our theoretical finding.}\label{tab: eg1}
\begin{tabular}{c|ccccc }
\multicolumn{1}{c}{} & \multicolumn{5}{c}{$\mbox{Sub-case 1: } e_1 = 0.2, e_2 =0.2$}    \\ 
\cmidrule(lr){2-6} 
Time Points & RE & $SN_{S_1}$ & $SP_{S_1}$ & $SN_{S_2}$ & $SP_{S_2}$  \\
\hspace{0.5 em} \\
\hline
100 & 0.09 & 0.93 & 0.82 & 1 & 0.89  \\
200 & 0.06 & 0.96 & 0.95 & 0.95 & 0.98  \\
300 & 0.06 & 0.98 & 0.97 & 0.95 & 1 \\
\hline
\end{tabular}

\bigskip

\begin{tabular}{c|ccccc }
\multicolumn{1}{c}{} & \multicolumn{5}{c}{$\mbox{Sub-case 2: } e_1 = 0.4, e_2 =0.4$}    \\ 
\cmidrule(lr){2-6}
Time Points & RE & $SN_{S_1}$ & $SP_{S_1}$ & $SN_{S_2}$ & $SP_{S_2}$  \\
\hspace{0.5 em} \\
\hline
100 & 0.17 & 0.82 & 0.78 & 0.88 & 0.83  \\
200 & 0.13 & 0.83 & 0.92 & 0.93 & 0.98  \\
300 & 0.09 & 0.90 & 0.90 & 0.93 & 1 \\
\hline
\end{tabular}
\end{table}

\begin{table}[h]
\caption{Performance Evaluation under setup 2: $d_1 = 30$, $d_2 = 20$. Relative error, sensitivity and specificity are reported for two different sparsity levels that is, edge densities 0.2 and 0.4. As the number of time points increases, estimation quality improves. Also, for any fixed time point, when the edge density increases from 0.2 to 0.4, the relative error increases, which is in line with our theoretical finding.}\label{tab: eg2}
\begin{tabular}{c|ccccc }
\multicolumn{1}{c}{} & \multicolumn{5}{c}{ $\mbox{Sub-case 1: }e_1 = 0.2, e_2 =0.2$}    \\ 
\cmidrule(lr){2-6} 
Time Points & RE & $SN_{S_1}$ & $SP_{S_1}$ & $SN_{S_2}$ & $SP_{S_2}$  \\
\hspace{0.5 em} \\
\hline
100 & 0.13 & 0.83 & 0.93 & 0.94 & 0.98  \\
200 & 0.08 & 0.93 & 0.91 & 0.94 & 1  \\
300 & 0.07 & 0.93 & 0.96 & 0.94 & 1 \\
\hline
\end{tabular}

\bigskip

\begin{tabular}{c|ccccc }
\multicolumn{1}{c}{} & \multicolumn{5}{c}{$ \mbox{Sub-case 2: }e_1 = 0.4, e_2 =0.4$}    \\ 
\cmidrule(lr){2-6}
Time Points & RE & $SN_{S_1}$ & $SP_{S_1}$ & $SN_{S_2}$ & $SP_{S_2}$  \\
\hspace{0.5 em} \\
\hline
100 & 0.22 & 0.82 & 0.74 & 0.84 & 0.97  \\
200 & 0.16 & 0.84 & 0.82 & 0.87 & 1  \\
300 & 0.11 & 0.88 & 0.91 & 0.91 & 1 \\
\hline
\end{tabular}
\end{table}

\begin{table}[h]
\caption{Performance Evaluation under setup 3: $d_1 = 15$, $d_2 = 10$. Relative error and the ranks of the estimated matrices are reported for different true rank values $R_1$ and $R_2$. As the number of time points increases, estimation quality improves. Also, for any fixed time point, when the true rank increases, the relative error increases, which aligns with our theoretical finding.}\label{tab:eg3}
\begin{tabular*}{\textwidth}{@{\extracolsep\fill}lcccccc}
\toprule%
& \multicolumn{3}{@{}c@{}}{$\mbox{Sub-case 1: }R_1=3, R_2=3$} & \multicolumn{3}{@{}c@{}}{$\mbox{Sub-case 2: }R_1=5, R_2=5$} \\\cmidrule{2-4}\cmidrule{5-7}%
Time points & RE & $\hat{R}_1$ & $\hat{R}_2$ & RE & $\hat{R}_1$ & $\hat{R}_2$ \\
\midrule
100  & 0.21  & 4 & 3 &0.23 & 5& 5\\
200  & 0.18  & 3 & 3 &0.18 & 5& 5\\
300  & 0.11  & 3 & 3 &0.15 & 5& 5\\
\botrule
\end{tabular*}
\end{table}

\begin{table}[h]
\caption{Performance Evaluation under setup 4: $d_1 = 30$, $d_2 = 20$. Relative error and the ranks of the estimated matrices are reported for different true rank values $R_1$ and $R_2$. As the number of time points increases, estimation quality improves. Also, for any fixed time point, when the true rank increases, the relative error increases, which aligns with our theoretical finding.}\label{tab:eg4}
\begin{tabular*}{\textwidth}{@{\extracolsep\fill}lcccccc}
\toprule%
& \multicolumn{3}{@{}c@{}}{$\mbox{Sub-case 1: }R_1=3, R_2=3$} & \multicolumn{3}{@{}c@{}}{$\mbox{Sub-case 2: }R_1=5, R_2=5$} \\\cmidrule{2-4}\cmidrule{5-7}%
Time points & RE & $\hat{R}_1$ & $\hat{R}_2$ & RE & $\hat{R}_1$ & $\hat{R}_2$ \\
\midrule
100  & 0.22  & 3 & 3 &0.24 & 5& 5\\
200  & 0.16  & 3 & 3 &0.18 & 5& 5\\
300  & 0.13  & 3 & 3 &0.15 & 5& 5\\
\botrule
\end{tabular*}
\end{table}

\noindent
\subsection{Predictive performance}\label{Sim:pred}
\noindent We now assess the predictive performance of our model and compare it against the bilinear MAR model \citep{chen2021autoregressive} and the sparse vector autoregressive model \citep{basu2015regularized}. The bilinear MAR model, as mentioned earlier in Section \ref{intro}, uses a multiplicative interaction of row-wise and column-wise temporal dependence with a bilinear form. On the other hand, to apply the sparse VAR model to our matrix-variate time series, we simply vectorize the matrix data, and apply sparsity regularization on that vector. We first fix a forecast horizon `h'. Then, for each $t^\prime \in \{ T-10, T-9 \dots , T-h\}$, we use all the data up to time point $t^\prime$ to estimate the model parameters, and finally we use that model to predict the value of $Y_{t^\prime+h}$, which is denoted by  $\hat{Y}_{t^\prime+h}$. Then, for that forecast horizon `h', the Root Mean Squared Error (RMSE) is defined as $ \sqrt{ \frac{1}{10-h+1} \sum_{t'=T-10}^{T-h} \frac{\left \lVert Y_{t'+h}- \hat{Y}_{t'+h}  \right \rVert_{F}^{2} }{d_1 d_2} }$, as in \cite{ghosh2018high} and \cite{chakraborty2023bayesian}. To examine the predictive performance of our model, we use a simulated data with $d_1 = 10, d_2 = 15$ and $T=80$. The true ranks of ${L_1}$ and ${L_2}$ are taken as 3 and 4 respectively, while the true edge densities of ${S_1}$ and ${S_2}$ are taken as 0.5 and 0.3 respectively. We consider forecast horizon values $h=1,2,3$ and compare the RMSE values of our model with that of the bilinear MAR and the sparse vector autoregressive model. As summarized in Table \ref{table-pred perf-simu}, RMSE values for our model are lower than both the bilinear MAR and the sparse VAR model, demonstrating better predictive performance of our model. As expected, the sparse VAR model exhibits poor predictive performance due to its naive vectorization of the matrix-variate time series, which disregards the inherent row-column interactions within the data. While the bilinear MAR model performs reasonably well in forecasting, the proposed additive MAR consistently outperforms it across all forecasting horizons, highlighting its superior predictive ability alongside other strengths of this model discussed earlier. 

\begin{table}[h]
\caption{Predictive performance using RMSE values. The proposed additive MAR model performs better than the competing bilinear MAR model and the sparse VAR model.}\label{table-pred perf-simu}%
\begin{tabular}{@{}llll@{}}
\toprule
Forecast horizon (h) & Additive MAR & Bilinear MAR & Sparse VAR\\
\midrule
 1   & 0.530   & 0.538   & 1.020   \\
 2   & 0.529   & 0.536   & 0.767   \\
 3   & 0.533   & 0.534   & 0.767 \\
\botrule
\end{tabular}
\end{table}
\noindent\textbf{AIC Criteria}\\

\noindent As mentioned earlier in this section, while working with real data, the true rank and the true sparsity levels are unknown. In such situations, we choose the values of $\lambda_{L_1}$, $\lambda_{S_1}$, $\lambda_{L_2}$ and $\lambda_{S_2}$ in such a way that the AIC, as defined below, is minimized.

$$
AIC = T\log\left(\frac{RSS}{T}\right)+ 2 \text{ Rank }(\hat{L}_1)+ 2 \text{ Rank }(\hat{L}_2) + 2k_1 + 2k_2
$$
where RSS, the residual sum of square, is defined as $\frac{1}{2T} \sum_{t=1}^{T} \left\lVert Y_t  - (L_1 + S_1)Y_{t-1} - Y_{t-1}(L_2+S_2)^\prime \right\rVert_{F}^{2}$, and $k_1$ and $k_2$ are the number of non-zero elements in $\hat{S}_1$ and $\hat{S}_2$ respectively. This formulation is quite common in the literature, which essentially rewards goodness of fit, and at the same time it 
penalizes overfitting.

\section{Application in Macroeconomic data}
\label{real_data}

We illustrate our proposed model using a matrix-valued time series data observed quarterly, from 2002-Q2 to 2019-Q4, comprising $16 (=d_1)$ key macroeconomic indicators for $11 (=d_2)$ Eurozone countries, namely, Austria, Belgium, Finland, France, Germany, Greece, Ireland, Italy, Netherlands, Portugal and Spain. Some necessary transformations, as suggested in \cite{mccracken2020fred} and \cite{stock2005empirical}, are applied to the macroeconomic variables in order to address the issue of non-stationarity. A summary of the macroeconomic variables along with the transformations applied on them and their sources are listed below in Table \ref{Analysis1}.

\begin{table}[h]
\caption{Details of the Macroeconomic Variables.}\label{Analysis1}%
\begin{tabular}{@{}llll@{}}
\toprule
Variable & Abbreviation & Source & Transformation \\
\midrule
 Interest Rate of Long-Term Government Bond Yields& GOV. BOND & EUROSTAT & $\Delta$\\
     Consumer Price Index: All Items& CPI & IMF & $\Delta^2 \ln$\\
     Producer Price Index: All Commodities & PPI & IMF & $\Delta^2 \ln$\\
     Total Share Prices for All Shares& Tot\textunderscore Share & FRED & $\Delta^2 \ln$\\
     Final Consumption Expenditure & Cons\textunderscore Exp&IMF& $\Delta \ln$ \\
     Capacity Utilization& Cap\textunderscore Util& FRED& $\Delta$\\
     All Employees & Empl & FRED & $\Delta \ln$\\
     Civilian Unemployment Rate & Un\textunderscore Rate & FRED & $\Delta$\\
     Compensation of Employees & Comp & IMF& $\Delta \ln$ \\
     National Income & Nat\textunderscore Income & IMF & $\Delta \ln$ \\
     Effective Exchange Rate (based on Unit-Labor-Cost )& EER & IMF&$\Delta$\\
     Industrial Production Index & IPI & IMF &$\Delta$\\
     Total Reserves &Tot\textunderscore Res&IMF& $\Delta^2 \ln$ \\
     External Balance of Goods and Services&BGS&IMF&$\Delta \ln$\\
     Broad Money Liabilities&M\textunderscore 2&IMF&$\Delta^2 \ln$\\
     Gross Domestic Product deflator &GDP&IMF&$\Delta^2 \ln$\\
\botrule
\end{tabular}
\end{table}

Following equation \eqref{main eqn 1}, we use $Y_t \in \mathbb{R}^{d_{1} \times d_{2}}$ to denote the matrix-valued observation at the $t^{th}$ quarter, whose $(i,j)^{th}$ element is the value corresponding to the $i^{th}$ macroeconomic variable for the $j^{th}$ country, $i= 1,2, \dots, d_1 = 16 ; j= 1,2, \dots, d_2 = 11$ and $t= 1,2, \dots, T = 71$. 

We first choose the regularization parameters $\lambda_{L_1}$, $\lambda_{L_2}$, $\lambda_{S_1}$, $\lambda_{S_2}$ using the AIC criteria discussed in Section \ref{simu}. The estimated component ${\hat{L}_1}$, as in equation \eqref{main eqn 2}, is a $16 \times 16$ matrix capturing the `baseline' component of the economic indicator-wise temporal dependence. The rank of $\hat{L}_1$ turns out to be $14$, indicating that the temporal dependence patterns of $14$ out of the $16$ economic indicators are mutually independent. However, the remaining two indicators exhibit temporal dependence patterns that can be expressed as linear combinations of those of the $14$ indicators. Upon further examination, we found that Broad Money Liabilities and Gross Domestic Product deflator are the two indicators whose temporal dependence patterns are linearly dependent on the others. Similarly, the estimated component ${\hat{L}_2}$ is a $11 \times 11$ matrix capturing the `baseline' component of the country-wise temporal dependence. The rank of $\hat{L}_2$ turns out to be 8, suggesting that the underlying temporal dependence patterns of 8 out of 11 countries are linearly independent, while the remaining three can be expressed as linear combinations of these. Further analysis reveals that the Netherlands, Portugal and Spain are the countries whose temporal dependence patterns are linearly dependent on those of the other eight countries.

\begin{figure}[H]
    \centering
    \includegraphics[scale=0.3]{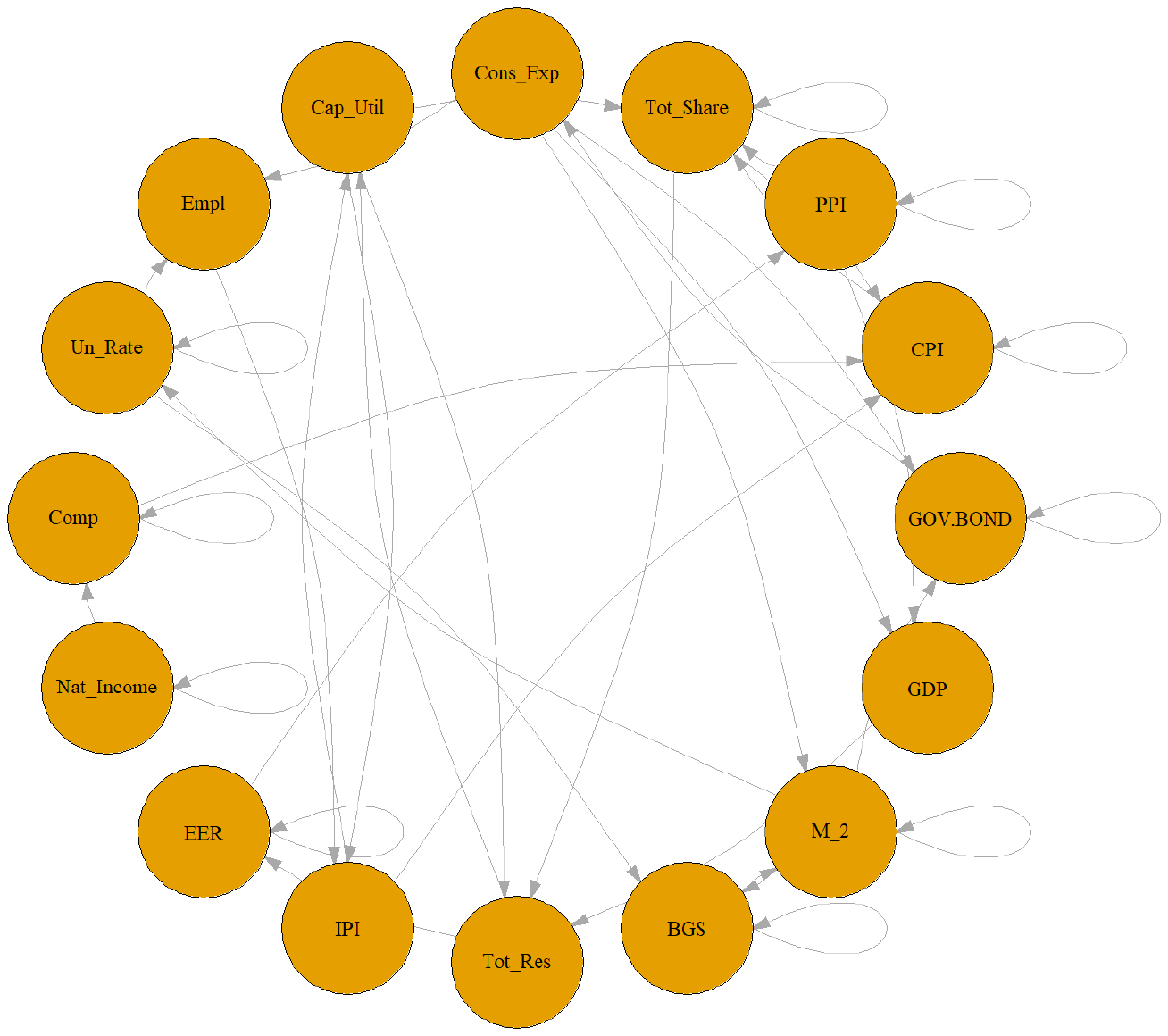}
    \caption{Network connectivity plot to represent the additional idiosyncratic temporal dependence among the 16 economic indicators, captured by the sparse component $\hat{S}_1$. Out of the total $16^2$ potential connections, only those with directed edges represent non-zero connections. For example, the directed edge from Effective Exchange Rate (EER) to Producer Price Index (PPI) represents one such indicator-wise temporal dependence.}  
    \label{circ connectivity indicator}
\end{figure}

\begin{figure}[H]
    \centering
    \includegraphics[scale =0.3]{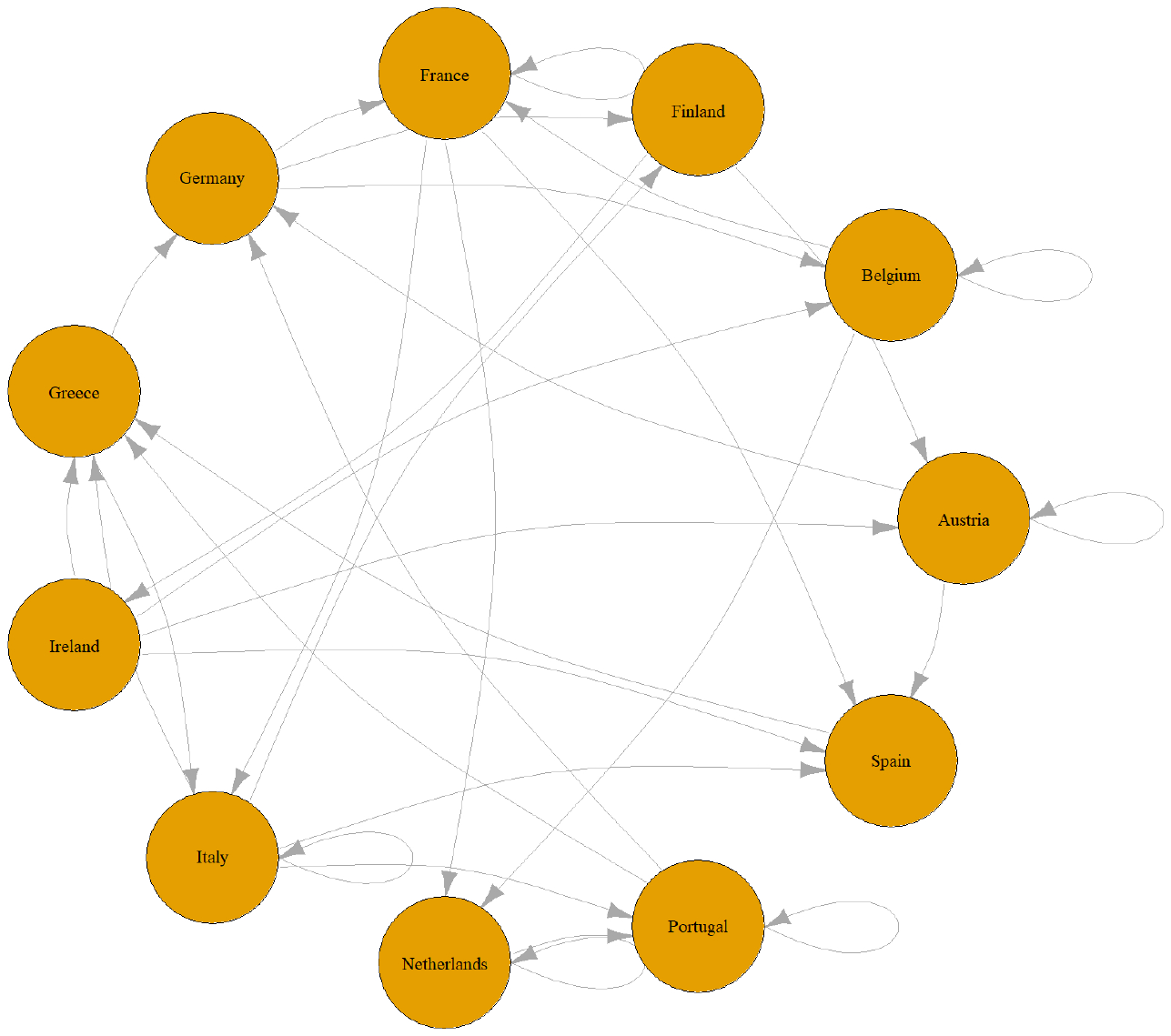}
    \caption{Network connectivity plot to represent the additional idiosyncratic temporal dependence among the 11 countries, captured by the sparse component $\hat{S}_2$. Out of the total $11^2$ potential connections, only those with directed edges represent non-zero connections. For example, the directed edge from France to Netherlands represents one such country-wise temporal dependence.}
    \label{circ connectivity country}
\end{figure}

The estimated sparse components ${\hat{S}_1}$ and ${\hat{S}_2}$ capture the additional idiosyncratic components of the economic indicator-wise temporal dependence and country-wise temporal dependence respectively. It turns out that $\hat{S}_1$ and $\hat{S}_2$ have edge densities $0.15$ and $0.26$ respectively. In other words, out of $16^2$ total indicator-wise temporal connections in $\hat{S}_1$, around 15\% are non-zero and the remaining are all zeros. Similarly, out of $11^2$ total country-wise temporal connections in $\hat{S}_2$, around 26\% are non-zero and the remaining are all zeros. These idiosyncratic temporal connections in $\hat{S}_1$ and $\hat{S}_2$, in addition to the aforementioned baseline temporal connections captured in $\hat{L}_1$ and $\hat{L}_2$, arise due to a period of financial crisis or economic boom in some specific countries, impacting the temporal relations between some specific economic indicators; for example, Greece Government
debt crisis, Portuguese financial crisis. We use two circular network graphs in Figure \ref{circ connectivity indicator} and Figure \ref{circ connectivity country} that illustrate the idiosyncratic temporal connections in ${\hat{S}_1}$ and ${\hat{S}_2}$ respectively, where each directed edge represents a non-zero temporal dependence. For example, as depicted in  Figure \ref{circ connectivity indicator}, the directed edge from Effective Exchange Rate (EER) to Producer Price Index (PPI) represents one such indicator-wise temporal dependence. Similarly, the directed edge from France to Netherlands in Figure \ref{circ connectivity country}  is one such country-wise temporal connection.    


Finally, we evaluate the predictive performance of our proposed regularized additive MAR model on this dataset and compare it with the competing bilinear MAR model and the sparse VAR model. As described earlier in Sections \ref{intro} and \ref{simu}, the bilinear MAR model uses a multiplicative interaction of row-wise and column-wise temporal dependence with a bilinear form. On the other hand, to apply the sparse VAR model to our matrix-variate time series, we simply vectorize the matrix data, and apply sparsity regularization on that vector. Table \ref{table-pred perf} summarizes the RMSE values, defined in Section \ref{simu}, for all the three models across the forecast horizons 1, 2 and 3. As the table illustrates, RMSE values are consistently lower for our model for all the forecast horizons, indicating improved predictive performance of our method as compared to the bilinear MAR model and the sparse VAR model. This aligns with our simulation results presented in Section \ref{simu}, where the sparse VAR model performs notably worse -- unsurprisingly, as it vectorizes matrix time series, thereby discarding important structural information intrinsic to the matrix form. While the bilinear MAR model performs better than the sparse VAR, it still consistently underperforms compared to our method, further validating the ability of our proposed approach in achieving improved forecasting accuracy.

\begin{table}[h]
\caption{Predictive performance using RMSE values. The proposed additive MAR model performs better than the competing bilinear MAR model and the sparse VAR model.}\label{table-pred perf}%
\begin{tabular}{@{}llll@{}}
\toprule
Forecast horizon (h) & Additive MAR & Bilinear MAR & Sparse VAR\\
\midrule
 1   & 0.761   & 0.801   & 1.229   \\
 2   & 0.747   & 0.794   & 1.113   \\
 3   & 0.746   & 0.798   & 1.053 \\
\botrule
\end{tabular}
\end{table}

\section{Discussion}
\label{disc}
In this work, we propose a high-dimensional regularized additive matrix autoregressive model that captures the temporal dependence among the matrix-valued time series by employing an additive interaction form, wherein the overall temporal connection is represented as the sum of row-wise and column-wise temporal dependence in the data. To accommodate high-dimensionality of the parameters, we then impose different regularized structures on row-wise and column-wise transition matrices $\textendash$ low-rank, sparse, or low-rank plus sparse decomposed structure, depending on the context. As discussed in \cite{zhang2024additive}, this additive interaction form, as opposed to convoluted bilinear representation, offers more comprehensible interpretation of the row-wise and column-wise temporal dependence. Also, with additive form, the penalized transition matrices help in extracting meaningful low-dimensional pattern in the data, whereas, the same with bilinear form provides only dimension reduction.

Some future research directions along this line are discussed next. First, this method can be readily extended to a three-dimensional or higher-order tensor setting \citep{kolda2009tensor}. For three-dimensional tensor-variate time series data, \cite{li2021multi} used a convoluted form of the temporal dependence using a Tucker-decomposed structure \citep{kolda2009tensor}. However, instead of that convoluted form, one can adopt a simple extension of our proposed additive interaction form in this paper $\textendash$ in the three-dimensional tensor case, the overall temporal dependence will be the addition of temporal dependence along the three modes of the tensor: row-wise temporal dependence, column-wise temporal dependence and tube-wise temporal dependence \citep{kolda2009tensor,roy2022regularized}. In terms of notation, this implies that the temporal dependence structure can be expressed as the sum of $Y_{t-1}\times_1A_1$, $Y_{t-1}\times_2A_2$ and $Y_{t-1}\times_3A_3$. Here, the transition matrices $A_1$, $A_2$ and $A_3$, that capture the row-wise, column-wise and tube-wise temporal dependence in the data respectively, are multiplied along the three modes of the tensor $Y_{t-1}$ using the mode-wise products $\times_1$, $\times_2$ and $\times_3$ respectively \citep{kolda2009tensor}. Thus, similar to the matrix case discussed in this paper, our method helps to disjoin and interpret temporal dependence across different modes in the tensor setting. Secondly, \cite{wang2019factor} proposed a factor model for matrix-variate time series, where they pre-multiplied and post-multiplied the core factor matrix $F_t$ with the front-loading (or, row-wise loading) $R$ and back-loading (or, column-wise loading) $C$ matrices respectively, yielding the bilinear form $RF_tC^\prime$. In contrast, it would be interesting to explore whether an additive row-wise and column-wise factor-loading representation can be employed by borrowing the idea from this paper. Finally, while the upper bound of the estimation error in this work has been derived under the assumption of Gaussian errors, it would be valuable to investigate how the upper bound generalizes under the sub-exponential distributional assumption of the errors.

\bibliography{sn-bibliography}

\begin{appendices}
    \section{Proofs of the theoretical results}
\subsection*{Basic Inequality}

    \begin{flalign}
         &\frac{1}{2T} \sum_{t=1}^{T} \left\lVert \big[ {\hat{\Delta}_{{L_{1}}} + \hat{\Delta}_{{S_{1}}}} \big] {Y_{t-1}} + {Y_{t-1}} \big[ {\hat{\Delta}_{{L_{2}}} + \hat{\Delta}_{{S_{2}}}} \big]^{T} \right\rVert _{F}^{2} \nonumber \\
          &\leq
         \frac{1}{T} \sum_{t=1}^{T} \biggl \langle {E_{t}}, \big[{\hat{\Delta}_{{L_{1}}} + \hat{\Delta}_{{S_{1}}}} \big] {Y_{t-1}} + {Y_{t-1}} \big[ {\hat{\Delta}_{{L_{2}}} + \hat{\Delta}_{{S_{2}}}}\big]^{T}\biggr \rangle + \lambda_{{L_{1}}} {C_{1}}({L_{1},S_{1}}) + \lambda_{{L_{2}}} {C_{2}}({L_{2},S_{2}}) \nonumber \\
         & - \lambda_{{L_{1}}} {C_{1}}({L_{1} +\hat{\Delta}_{{L_{1}}},S_{1}+\hat{\Delta}_{{S_{1}}}}) - \lambda_{{L_{2}}} {C_{2}}({L_{2} +\hat{\Delta}_{{L_{2}}},S_{2}+\hat{\Delta}_{{S_{2}}}})
         \label{eq41}
      \end{flalign}

\textit{Proof}: We may note that the following inequality holds from the optimality of $({\hat{L}_{1}, \hat{L}_{2}, \hat{S}_{1}, \hat{S}_{2}})$ and the feasibility of $({{L}_{1}, {L}_{2}, {S}_{1}, {S}_{2}})$. \\

\begin{flalign}
    &\frac{1}{2T} \sum_{t=1}^{T} \left \lVert {Y_{t} - (\hat{L}_{1} +\hat{S}_{1} )Y_{t-1} - Y_{t-1}(\hat{L}_{2} +\hat{S}_{2})^{T}} \right \lVert_{F}^{2} + \lambda_{{S_{1}}} \left \lVert {\hat{S}_{1}} \right \rVert _{1} + \lambda_{{S_{2}}} \left \lVert {\hat{S}_{2}} \right \rVert _{1} 
     + \lambda_{{L_{1}}} \left \lVert {\hat{L}_{1}} \right \rVert _{*} + 
     \lambda_{{L_{2}}} \left \lVert {\hat{L}_{2}} \right \rVert _{*}\nonumber \\ &\leq 
     \frac{1}{2T} \sum_{t=1}^{T} \left \lVert {Y_{t} - (L_{1} +S_{1} )Y_{t-1} - Y_{t-1}(L_{2} +S_{2})^{T}} \right \lVert_{F}^{2} + \lambda_{{S_{1}}} \left \lVert {S_{1}} \right \rVert _{1} + \lambda_{{S_{2}}} \left \lVert {S_{2}} \right \rVert _{1} + \lambda_{{L_{1}}} \left \lVert {L_{1}} \right \rVert _{*} \nonumber \\
    & + \lambda_{{L_{2}}} \left \lVert {L_{2}} \right \rVert _{*} \label{eq8}
\end{flalign}

Now, from our model, ${Y_{t}}= ({L_{1}} + {S_{1}}){Y_{t-1}} + {Y_{t-1} }({L_{2}+S_{2}})^{T} + {E_{t}}$, we have the following,

\begin{align}
    &\sum_{t=1}^{T} \left \lVert {Y_{t}} - \Big[({\hat{L_{1}}} + {\hat{S_{1}}}){Y_{t-1}} + {Y_{t-1} }({\hat{L_{2}}+\hat{S_{2}}})^{T} \Big] \right\rVert_{F}^{2} \nonumber \\
        &= \sum_{t=1}^{T} \left\lVert {E_{t}} - \Big[({\hat{L_{1}}} + {\hat{S_{1}}}){Y_{t-1}} + {Y_{t-1} }({\hat{L_{2}}+\hat{S_{2}}})^{T} \Big] + \Big[({L_{1}} + {S_{1}}){Y_{t-1}} + {Y_{t-1} }({L_{2}+S_{2}})^{T} \Big] \right\rVert_{F}^{2} \nonumber \\
        &= \sum_{t=1}^{T} \left\lVert {E_{t}}- \Big[ ({\hat{{L}_{1}}} - {L_{1}}) + ({\hat{{S}_{1}}} - {S_{1}})\Big]{Y_{t-1}} - {Y_{t-1}}\Big[ ({\hat{{L}_{2}}}^{T} - {L_{2}}^{T}) + ({\hat{{S}_{2}}}^{T} - {S_{2}}^{T})\Big]\right\rVert_{F}^{2} \nonumber \\
        &= \sum_{t=1}^{T} \left\lVert {E_{t}} - \Big[ {\hat{\Delta}_{L_{1}}} + {\hat{\Delta}_{S_{1}}}\Big]{Y_{t-1}}-{Y_{t-1}}\Big[{\hat{\Delta}_{L_{2}}} + {\hat{\Delta}_{S_{2}}}\Big]^{T}\right\rVert_{F}^{2} \label{eq99} \\
        \nonumber
\end{align} 
 Now, let us define, $${B_{t}} = \Big[ {\hat{\Delta}_{L_{1}}} + {\hat{\Delta}_{S_{1}}}\Big]{Y_{t-1}}+{Y_{t-1}}\Big[{\hat{\Delta}_{L_{2}}} + {\hat{\Delta}_{S_{2}}}\Big]^{T}$$ 
        Then, the quantity in \ref{eq99} reduces to the following:
        \begin{align}
            &\sum_{t=1}^{T} \left\lVert {E_{t}} - {B_{t}} \right\rVert _{F}^{2} \nonumber \\
            &=  \sum_{t=1}^{T} \left\lVert {E_{t}}\right\rVert _{F}^{2} +  \sum_{t=1}^{T} \left\lVert {B_{t}}\right\rVert _{F}^{2} - 2 \sum_{t=1}^{T} \biggl \langle {E_{t}}, {B_{t}}\biggr \rangle \label{eq10}
        \end{align}
   
Now, we combine the decomposition in  \ref{eq10} with the inequality in \ref{eq8} to arrive at the following proof of this lemma.
\begin{flalign}
    & \frac{1}{2T} \sum_{t=1}^{T}  \left\lVert {E_{t}}\right\rVert_{F}^{2} + \frac{1}{2T} \sum_{t=1}^{T}  \left\lVert {B_{t}}\right\rVert_{F}^{2} - \frac{1}{T} \sum_{t=1}^{T} \biggl \langle {E_{t}}, {B_{t}}\biggr \rangle + \lambda_{{S_{1}}} \left\lVert {\hat{S}_{1}} \right\rVert_{1} + \lambda_{{S_{2}}} \left\lVert {\hat{S}_{2}} \right\rVert_{1} + \lambda_{{L_{1}}} \left\lVert {\hat{L}_{1}} \right\rVert_{*} \nonumber \\
    &+ \lambda_{{L_{2}}} \left\lVert {\hat{L}_{2}} \right\rVert_{*} \leq \frac{1}{2T} \sum_{t=1}^{T}  \left\lVert {E_{t}}\right\rVert_{F}^{2} + \lambda_{{S_{1}}} \left\lVert {S_{1}} \right\rVert_{1} + \lambda_{{S_{2}}} \left\lVert {S_{2}} \right\rVert_{1} +  \lambda_{{L_{1}}} \left\lVert {L}_{1} \right\rVert_{*} + \lambda_{{L_{2}}} \left\lVert {L}_{2} \right\rVert_{*}
\end{flalign}

\subsection*{\textit{Proof of Lemma \ref{lemma:31}}}

Using Asusmption \ref{ass 1}, we get the following, 
\begin{flalign}
    &\frac{1}{2T} \sum_{t=1}^{T} \left\lVert \big[ {\hat{\Delta}_{{L_{1}}} + \hat{\Delta}_{{S_{1}}}} \big] {Y_{t-1}} + {Y_{t-1}} \big[ {\hat{\Delta}_{{L_{2}}} + \hat{\Delta}_{{S_{2}}}} \big]^{T} \right\rVert _{F}^{2} \nonumber \\
    \label{eq9}
    & \geq \frac{\gamma}{2} \Bigg[ \left \lVert {\hat{\Delta}_{{L_{1}}}} + {\hat{\Delta}_{{S_{1}}}}  \right \rVert ^{2}_{F} + \left \lVert {\hat{\Delta}_{{L_{2}}}} + {\hat{\Delta}_{{S_{2}}}}  \right \rVert ^{2}_{F}\Bigg]
\end{flalign}

We intend to find a lower bound for the right-hand side of the above inequality and an upper bound for the left-hand side of the same. We start with the derivation of the lower bound for $\Bigg[ \left \lVert {\hat{\Delta}_{{L_{1}}}} + {\hat{\Delta}_{{S_{1}}}}  \right \rVert ^{2}_{F} + \left \lVert {\hat{\Delta}_{{L_{2}}}} + {\hat{\Delta}_{{S_{2}}}}  \right \rVert ^{2}_{F}\Bigg]$. One may note that,

\begin{flalign}
    &\frac{\gamma}{2} \Bigg[ \left \lVert {\hat{\Delta}_{{L_{1}}}} + {\hat{\Delta}_{{S_{1}}}}  \right \rVert ^{2}_{F} + \left \lVert {\hat{\Delta}_{{L_{2}}}} + {\hat{\Delta}_{{S_{2}}}}  \right \rVert ^{2}_{F}\Bigg] \nonumber \\ 
    &= \frac{\gamma}{2} \Bigg[ \left \lVert  {\hat{\Delta}_{{L_{1}}}}\right\rVert _F^{2} + \left \lVert {\hat{\Delta}_{{S_{1}}}}\right\rVert _{F}^{2} + 2 \biggl \langle  {\hat{\Delta}_{{L_{1}}}}, {\hat{\Delta}_{{S_{1}}}} \biggr\rangle  + \left \lVert  {\hat{\Delta}_{{L_{2}}}}\right\rVert _{F}^{2} + \left \lVert {\hat{\Delta}_{{S_{2}}}}\right\rVert _{F}^{2} + 2 \biggl \langle  {\hat{\Delta}_{{L_{2}}}}, {\hat{\Delta}_{{S_{2}}}} \biggr\rangle\Bigg]
\end{flalign}

From the above equation, we obtain the following,

\begin{flalign}
    &\frac{\gamma}{2} \Bigg[ \left \lVert {\hat{\Delta}_{{L_{1}}}}\right\rVert _{F}^{2} +  \left \lVert {\hat{\Delta}_{{S_{1}}}}\right\rVert _{F}^{2} +  \left \lVert {\hat{\Delta}_{{L_{2}}}}\right\rVert _{F}^{2} +  \left \lVert {\hat{\Delta}_{{S_{2}}}}\right\rVert _{F}^{2} \Bigg] -
    \frac{\gamma}{2} \Bigg[\left \lVert {\hat{\Delta}_{{L_{1}}}} + {\hat{\Delta}_{{S_{1}}}}  \right \rVert ^{2}_{F} + \left \lVert {\hat{\Delta}_{{L_{2}}}} + {\hat{\Delta}_{{S_{2}}}}  \right \rVert ^{2}_{F} \Bigg] \nonumber \\ \label{myeq}
    &= -\gamma \Bigg[ \biggl \langle {\hat{\Delta}_{{L_{1}}}}, {\hat{\Delta}_{{S_{1}}}} \biggr \rangle + \biggl \langle {\hat{\Delta}_{{L_{2}}}}, {\hat{\Delta}_{{S_{2}}}} \biggr \rangle \Bigg] 
\end{flalign}

Using the Dual norm inequality, we may write,

\begin{flalign}
    \gamma \Big| \biggl\langle {\hat{\Delta}_{{L_{1}}}}, {\hat{\Delta}_{{S_{1}}}}   \biggr\rangle\Big| \leq \gamma \left \lVert {\hat{\Delta}_{{L_{1}}}}\right\rVert_{\infty} \left \lVert {\hat{\Delta}_{{S_{1}}}} \right\rVert_{1} \nonumber \\
    \gamma \Big| \biggl\langle {\hat{\Delta}_{{L_{2}}}}, {\hat{\Delta}_{{S_{2}}}}   \biggr\rangle\Big| \leq \gamma \left \lVert {\hat{\Delta}_{{L_{2}}}}\right\rVert_{\infty} \left \lVert {\hat{\Delta}_{{S_{2}}}} \right\rVert_{1}
\end{flalign}

\begin{flalign}
   & \Rightarrow \gamma \Bigg|\biggl\langle {\hat{\Delta}_{{L_{1}}}}, {\hat{\Delta}_{{S_{1}}}}   \biggr\rangle + \biggl\langle {\hat{\Delta}_{{L_{2}}}}, {\hat{\Delta}_{{S_{2}}}}   \biggr\rangle \Bigg| \leq \gamma \Bigg[\left \lVert {\hat{\Delta}_{{L_{1}}}}\right\rVert_{\infty} \left \lVert {\hat{\Delta}_{{S_{1}}}}\right\rVert_{1} + \left \lVert {\hat{\Delta}_{{L_{2}}}}\right\rVert_{\infty} \left \lVert {\hat{\Delta}_{{S_{2}}}}\right\rVert_{1} \Bigg] \nonumber \\
   & \leq \gamma \Bigg[ \bigg( \left \lVert {{{\hat{L_{1}}}}}\right\rVert_{\infty}  + \left \lVert {{{L_{1}}}}\right\rVert_{\infty} \bigg) \left \lVert {\hat{\Delta}_{{S_{1}}}}\right\rVert_{1} + \bigg( \left \lVert {{{\hat{L_{2}}}}}\right\rVert_{\infty}  + \left \lVert {{{L_{2}}}}\right\rVert_{\infty} \bigg) \left \lVert {\hat{\Delta}_{{S_{2}}}}\right\rVert_{1} \Bigg] \nonumber \\
   & \leq \gamma \Bigg[\frac{2 \alpha_1}{\sqrt{d_{1} d_1}} \left \lVert{\hat{\Delta}_{{S_{1}}}}\right\rVert_{1} + \frac{2 \alpha_2}{\sqrt{d_{2} d_2}} \left \lVert{\hat{\Delta}_{{S_{2}}}}\right\rVert_{1} \Bigg] \nonumber
\end{flalign}

Using equation \ref{myeq}, we arrive at the following inequality,

\begin{flalign}
    & \frac{\gamma}{2} \Bigg[ \left \lVert {\hat{\Delta}_{{L_{1}}}} + {\hat{\Delta}_{{S_{1}}}}  \right \rVert ^{2}_{F} + \left \lVert {\hat{\Delta}_{{L_{2}}}} + {\hat{\Delta}_{{S_{2}}}}  \right \rVert ^{2}_{F}\Bigg]  \nonumber \\
    & \geq \frac{\gamma}{2} \Bigg[\left \lVert {\hat{\Delta}_{{L_{1}}}}\right\rVert _{F}^{2} +  \left \lVert {\hat{\Delta}_{{S_{1}}}}\right\rVert _{F}^{2} +  \left \lVert {\hat{\Delta}_{{L_{2}}}}\right\rVert _{F}^{2} +  \left \lVert {\hat{\Delta}_{{S_{2}}}}\right\rVert _{F}^{2}\Bigg] - \gamma \Bigg[\frac{2 \alpha_1}{\sqrt{d_{1} d_1}} \left \lVert{\hat{\Delta}_{{S_{1}}}}\right\rVert_{1} + \frac{2 \alpha_2}{\sqrt{d_{2} d_2}} \left \lVert{\hat{\Delta}_{{S_{2}}}}\right\rVert_{1} \Bigg] \nonumber \\
    & \geq \frac{\gamma}{2} \Bigg[\left \lVert {\hat{\Delta}_{{L_{1}}}}\right\rVert _{F}^{2} +  \left \lVert {\hat{\Delta}_{{S_{1}}}}\right\rVert _{F}^{2} +  \left \lVert {\hat{\Delta}_{{L_{2}}}}\right\rVert _{F}^{2} +  \left \lVert {\hat{\Delta}_{{S_{2}}}}\right\rVert _{F}^{2} \Bigg] -\frac{\lambda_{{S_1}}}{2} \left \lVert {\hat{\Delta}_{{S_{1}}}}\right\rVert_{1} - \frac{\lambda_{{S_2}}}{2} \left \lVert {\hat{\Delta}_{{S_{2}}}}\right\rVert_{1} \nonumber \\
    & \geq \frac{\gamma}{2} \Bigg[\left \lVert {\hat{\Delta}_{{L_{1}}}}\right\rVert _{F}^{2} +  \left \lVert {\hat{\Delta}_{{S_{1}}}}\right\rVert _{F}^{2} +  \left \lVert {\hat{\Delta}_{{L_{2}}}}\right\rVert _{F}^{2} +  \left \lVert {\hat{\Delta}_{{S_{2}}}}\right\rVert _{F}^{2} \Bigg] -\frac{\lambda_{{S_1}}}{2} \left \lVert {\hat{\Delta}_{{S_{1}}}}\right\rVert_{1} \nonumber  - \frac{\lambda_{{S_2}}}{2} \left \lVert {\hat{\Delta}_{{S_{2}}}}\right\rVert_{1} \\ &-  \frac{\lambda_{{L_1}}}{2} \left \lVert {\hat{\Delta}_{{L_{1}}}}\right\rVert_{*} - \frac{\lambda_{{L_2}}}{2} \left \lVert {\hat{\Delta}_{{L_{2}}}}\right\rVert_{*} \nonumber \\
    & \geq \frac{\gamma}{2} \Bigg[\left \lVert {\hat{\Delta}_{{L_{1}}}}\right\rVert _{F}^{2} +  \left \lVert {\hat{\Delta}_{{S_{1}}}}\right\rVert _{F}^{2} +  \left \lVert {\hat{\Delta}_{{L_{2}}}}\right\rVert _{F}^{2} +  \left \lVert {\hat{\Delta}_{{S_{2}}}}\right\rVert _{F}^{2} \Bigg] - \frac{\lambda_{{L_1}}}{2} {C_1(\hat{\Delta}_{{L_{1}}}, \hat{\Delta}_{{S_{1}}})} - \frac{\lambda_{{L_2}}}{2} {C_2(\hat{\Delta}_{{L_{2}}}, \hat{\Delta}_{{S_{2}}})} \label{eq13}
\end{flalign}
Now, we derive an upper bound for the left hand side of the inequality \ref{eq9}.\\  

Using the inequalities \ref{eq2}, \ref{eq3}  and \ref{eq41}, we arrive at the following inequality:

\begin{flalign}
    &\frac{1}{2T} \sum_{t=1}^{T} \left\lVert \big[ {\hat{\Delta}_{{L_{1}}} + \hat{\Delta}_{{S_{1}}}} \big] {Y_{t-1}} + {Y_{t-1}} \big[ {\hat{\Delta}_{{L_{2}}} + \hat{\Delta}_{{S_{2}}}} \big]^{T} \right\rVert _{F}^{2} \nonumber \\
    \label{eq14}
    &\leq
    \frac{1}{T} \sum_{t=1}^{T} \biggl \langle {E_{t}}, \big[{\hat{\Delta}_{{L_{1}}} + \hat{\Delta}_{{S_{1}}}} \big] {Y_{t-1}} + {Y_{t-1}} \big[ {\hat{\Delta}_{{L_{2}}} + \hat{\Delta}_{{S_{2}}}}\big]^{T}\biggr \rangle + \lambda_{{L_{1}}} \Big[{C_1}({\hat{\Delta}_{{L_{1}}}}^{A_1}, {\hat{\Delta}_{{S_{1}}}}^\mathbb{M}) - {C_1}({\hat{\Delta}_{{L_{1}}}}^{B_1}, {\hat{\Delta}_{{S_{1}}}}^\mathbb{M^{\bot}}) \Big] \\ \nonumber
    & +   \lambda_{{L_{2}}} \Big[{C_2}({\hat{\Delta}_{{L_{2}}}}^{A_2}, {\hat{\Delta}_{{S_{2}}}}^\mathbb{N}) - {C_2}({\hat{\Delta}_{{L_{2}}}}^{B_2}, {\hat{\Delta}_{{S_{2}}}}^\mathbb{N^{\bot}}) \Big] 
\end{flalign}
Now, we may note that,
\begin{align*}
    &\frac{1}{T} \sum_{t=1}^{T} \biggl \langle {E_{t}}, \big[{\hat{\Delta}_{{L_{1}}} + \hat{\Delta}_{{S_{1}}}} \big] {Y_{t-1}} + {Y_{t-1}} \big[ {\hat{\Delta}_{{L_{2}}} + \hat{\Delta}_{{S_{2}}}}\big]^{T}\biggr \rangle \\ \nonumber
    & = \frac{1}{T} \sum_{t=1}^{T} \biggl \langle {E_{t}}, \big[{\hat{\Delta}_{{L_{1}}} + \hat{\Delta}_{{S_{1}}}} \big] {Y_{t-1}}\biggr \rangle + \frac{1}{T} \sum_{t=1}^{T} \biggl \langle{E_{t}}, {Y_{t-1}} \big[ {\hat{\Delta}_{{L_{2}}} + \hat{\Delta}_{{S_{2}}}}\big]^{T} \biggr \rangle \\ \nonumber
\end{align*}
Now, we may write the following,
\begin{align*}
    & \frac{1}{T} \sum_{t=1}^{T} \biggl \langle {E_{t}}, \big[{\hat{\Delta}_{{L_{1}}} + \hat{\Delta}_{{S_{1}}}} \big] {Y_{t-1}}\biggr \rangle \\
    & = \frac{1}{T} \sum_{t=1}^{T} \biggl \langle {E_{t}}{Y_{t-1}}^T, \big[{\hat{\Delta}_{{L_{1}}} + \hat{\Delta}_{{S_{1}}}} \big] \biggr \rangle \\
    & = \biggl \langle {\mathcal{D}_{1}}, \big[{\hat{\Delta}_{{L_{1}}} + \hat{\Delta}_{{S_{1}}}} \big]   \biggr \rangle \hspace{0.7 cm} \text{where,} \hspace{0.1cm}  {\mathcal{D}_{1}} =  \frac{1}{T} \sum_{t=1}^{T}  {E_{t}}{Y_{t-1}}^T \\
    & \leq
    \left \lVert {\hat{\Delta}_{{L_{1}}}} \right\rVert_{*}  \left \lVert {\mathcal{D}_1} \right\rVert_{sp} + \left \lVert {\hat{\Delta}_{{S_{1}}}} \right\rVert_{1}  \left \lVert {\mathcal{D}_1}
    \right\rVert_{\infty} \\
    & \leq
     \left \lVert {\mathcal{D}_1} \right\rVert_{sp} \Bigg[  \left \lVert {\hat{\Delta}_{{L_{1}}}}^{A_1} \right\rVert_{*} + \left \lVert {\hat{\Delta}_{{L_{1}}}}^{B_1} \right\rVert_{*}\Bigg] + 
     \left \lVert {\mathcal{D}_1} \right\rVert_{\infty} \Bigg[  \left \lVert {\hat{\Delta}_{{S_{1}}}}^{M} \right\rVert_{1} + \left \lVert {\hat{\Delta}_{{S_{1}}}}^{M^{\bot}} \right\rVert_{1}\Bigg]
\end{align*}

Using the definitions of ${C_1}({L_1} , {S_1})$ and ${C_2}({L_2} , {S_2})$ in \ref{defn C1 C2} and the assumptions on the regularization parameters in Assumption \ref{ass 3}, we get the following result,

\begin{align*}
    & \frac{1}{T} \sum_{t=1}^{T} \biggl \langle {E_{t}}, \big[{\hat{\Delta}_{{L_{1}}} + \hat{\Delta}_{{S_{1}}}} \big] {Y_{t-1}}\biggr \rangle 
    \leq 
    \frac{\lambda_{{L_1}}}{4}
    \Bigg[ {C_1}\big({\hat{\Delta}_{{L_{1}}}}^{A_1} + {\hat{\Delta}_{{S_{1}}}}^{M} \big) + {C_1}\big({\hat{\Delta}_{{L_{1}}}}^{B_1} + {\hat{\Delta}_{{S_{1}}}}^{M^{\bot}} \big)\Bigg]
\end{align*}

In the similar manner, we can show that,
\begin{align*}
    & \frac{1}{T} \sum_{t=1}^{T} \biggl \langle{E_{t}}, {Y_{t-1}} \big[ {\hat{\Delta}_{{L_{2}}} + \hat{\Delta}_{{S_{2}}}}\big]^{T} \biggr \rangle
    \leq
    \frac{\lambda_{{L_2}}}{4} \Bigg[{C_2}\big({\hat{\Delta}_{{L_{2}}}}^{A_2} + {\hat{\Delta}_{{S_{2}}}}^{N} \big) + {C_2}\big({\hat{\Delta}_{{L_{2}}}}^{B_2} + {\hat{\Delta}_{{S_{2}}}}^{N^{\bot}} \big) \Bigg]
\end{align*}
Using these two inequalities and \ref{eq14}, we can write the following,
\begin{align*}
    & \frac{1}{2T} \sum_{t=1}^{T} \left\lVert \big[ {\hat{\Delta}_{{L_{1}}} + \hat{\Delta}_{{S_{1}}}} \big] {Y_{t-1}} + {Y_{t-1}} \big[ {\hat{\Delta}_{{L_{2}}} + \hat{\Delta}_{{S_{2}}}} \big]^{T} \right\rVert _{F}^{2} \nonumber 
    \leq
     \frac{\lambda_{{L_1}}}{4}
    \Bigg[ {C_1}\big({\hat{\Delta}_{{L_{1}}}}^{A_1} + {\hat{\Delta}_{{S_{1}}}}^{M} \big) + {C_1}\big({\hat{\Delta}_{{L_{1}}}}^{B_1} + {\hat{\Delta}_{{S_{1}}}}^{M^{\bot}} \big)\Bigg] \\
    & + 
     \frac{\lambda_{{L_2}}}{4} \Bigg[{C_2}\big({\hat{\Delta}_{{L_{2}}}}^{A_2} + {\hat{\Delta}_{{S_{2}}}}^{N} \big) + {C_2}\big({\hat{\Delta}_{{L_{2}}}}^{B_2} + {\hat{\Delta}_{{S_{2}}}}^{N^{\bot}} \big) \Bigg]  + \lambda_{{L_{1}}} \Big[{C_1}({\hat{\Delta}_{{L_{1}}}}^{A_1}, {\hat{\Delta}_{{S_{1}}}}^\mathbb{M}) - {C_1}({\hat{\Delta}_{{L_{1}}}}^{B_1}, {\hat{\Delta}_{{S_{1}}}}^\mathbb{M^{\bot}}) \Big] \\ \nonumber
    & +   \lambda_{{L_{2}}} \Big[{C_2}({\hat{\Delta}_{{L_{2}}}}^{A_2}, {\hat{\Delta}_{{S_{2}}}}^\mathbb{N}) - {C_2}({\hat{\Delta}_{{L_{2}}}}^{B_2}, {\hat{\Delta}_{{S_{2}}}}^\mathbb{N^{\bot}}) \Big] 
\end{align*}
This reduces to the following,
\begin{align}
    & \frac{1}{2T} \sum_{t=1}^{T} \left\lVert \big[ {\hat{\Delta}_{{L_{1}}} + \hat{\Delta}_{{S_{1}}}} \big] {Y_{t-1}} + {Y_{t-1}} \big[ {\hat{\Delta}_{{L_{2}}} + \hat{\Delta}_{{S_{2}}}} \big]^{T} \right\rVert _{F}^{2}
    \leq
    \frac{3}{2} \lambda_{{L_1}} {C_1}({\hat{\Delta}_{{L_{1}}}}^{A_1}, {\hat{\Delta}_{{S_{1}}}}^\mathbb{M}) +
    \frac{3}{2} \lambda_{{L_2}} {C_2}({\hat{\Delta}_{{L_{2}}}}^{A_2}, {\hat{\Delta}_{{S_{2}}}}^\mathbb{N})\label{eq15}
\end{align}
Combining the inequalities in \ref{eq9}, \ref{eq13} and \ref{eq15}, we arrive at the 
following inequality,
\begin{align}
   & \frac{\gamma}{2} \Bigg[\left \lVert {\hat{\Delta}_{{L_{1}}}}\right\rVert _{F}^{2} +  \left \lVert {\hat{\Delta}_{{S_{1}}}}\right\rVert _{F}^{2} +  \left \lVert {\hat{\Delta}_{{L_{2}}}}\right\rVert _{F}^{2} +  \left \lVert {\hat{\Delta}_{{S_{2}}}}\right\rVert _{F}^{2}\Bigg] 
   \leq 
   \frac{3}{2} \lambda_{{L_1}} {C_1}({\hat{\Delta}_{{L_{1}}}}^{A_1}, {\hat{\Delta}_{{S_{1}}}}^\mathbb{M}) +
    \frac{3}{2} \lambda_{{L_2}} {C_2}({\hat{\Delta}_{{L_{2}}}}^{A_2}, {\hat{\Delta}_{{S_{2}}}}^\mathbb{N}) \label{eq16} \\ 
    & + \frac{\lambda_{{L_1}}}{2} {C_1(\hat{\Delta}_{{L_{1}}}, \hat{\Delta}_{{S_{1}}})} +\frac{\lambda_{{L_2}}}{2} {C_2(\hat{\Delta}_{{L_{2}}}, \hat{\Delta}_{{S_{2}}})} \nonumber 
\end{align}

We have the following results:

\begin{align}
    &{C_1}({\hat{\Delta}_{{L_{1}}}}, {\hat{\Delta}_{{S_{1}}}})
    \leq
    {C_1}({\hat{\Delta}_{{L_{1}}}}^{A_1}, {\hat{\Delta}_{{S_{1}}}}^\mathbb{M}) + {C_1}({\hat{\Delta}_{{L_{1}}}}^{B_1}, {\hat{\Delta}_{{S_{1}}}}^\mathbb{M^{\bot}}) \nonumber \\
    & {C_2}({\hat{\Delta}_{{L_{2}}}}, {\hat{\Delta}_{{S_{2}}}})
    \leq
    {C_2}({\hat{\Delta}_{{L_{2}}}}^{A_2}, {\hat{\Delta}_{{S_{2}}}}^\mathbb{N}) + {C_2}({\hat{\Delta}_{{L_{2}}}}^{B_2}, {\hat{\Delta}_{{S_{2}}}}^\mathbb{N^{\bot}})    
\end{align}

Combining these results with that of Lemma \ref{lemma:32}, we get the following,

\begin{align}
    & {C_1}({\hat{\Delta}_{{L_{1}}}}, {\hat{\Delta}_{{S_{1}}}})
    \leq
    4 {C_1}({\hat{\Delta}_{{L_{1}}}}^{A_1}, {\hat{\Delta}_{{S_{1}}}}^\mathbb{M}) \nonumber \\
    & {C_2}({\hat{\Delta}_{{L_{2}}}}, {\hat{\Delta}_{{S_{2}}}})
    \leq
    4 {C_2}({\hat{\Delta}_{{L_{2}}}}^{A_2}, {\hat{\Delta}_{{S_{2}}}}^\mathbb{N})
\end{align}

Using these results, we may rewrite \ref{eq16} in the following manner:

\begin{align}
    & \frac{\gamma}{2} \Bigg[\left \lVert {\hat{\Delta}_{{L_{1}}}}\right\rVert _{F}^{2} +  \left \lVert {\hat{\Delta}_{{S_{1}}}}\right\rVert _{F}^{2} +  \left \lVert {\hat{\Delta}_{{L_{2}}}}\right\rVert _{F}^{2} +  \left \lVert {\hat{\Delta}_{{S_{2}}}}\right\rVert _{F}^{2}\Bigg] 
   \leq \label{eq22}
   4 \lambda_{{L_1}} {C_1}({\hat{\Delta}_{{L_{1}}}}^{A_1}, {\hat{\Delta}_{{S_{1}}}}^\mathbb{M})
   + 
   4 \lambda_{{L_2}} {C_2}({\hat{\Delta}_{{L_{2}}}}^{A_2}, {\hat{\Delta}_{{S_{2}}}}^\mathbb{N})
\end{align}

We know from Lemma \ref{lemma:31}, that the rank of ${\hat{\Delta}_{{L_{1}}}}^{A_1}$ is at most 2$R_1$ and that of ${\hat{\Delta}_{{L_{2}}}}^{A_2}$ is at most 2$R_2$. We use this fact alongside the notion of \textit{Compatibility Constant} defined in \cite{r1} to arrive at the following inequalities,

\begin{align}
    & \lambda_{{L_1}}{C_1}({\hat{\Delta}_{{L_{1}}}}^{A_1}, {\hat{\Delta}_{{S_{1}}}}^\mathbb{M}) \leq \sqrt{2 R_1} \lambda_{{L_1}} \left \lVert {\hat{\Delta}_{{L_{1}}}}^{A_1} \right\rVert_{F} + \sqrt{s_1} \lambda_{{S_1}}\left \lVert {\hat{\Delta}_{{S_{1}}}}^{M} \right\rVert_{F} \nonumber \\ 
    & \hspace{3.3cm} \leq \sqrt{2 R_1} \lambda_{{L_1}}  \left \lVert {\hat{\Delta}_{{L_{1}}}} \right\rVert_{F} + \sqrt{s_1} \lambda_{{S_1}} \left \lVert {\hat{\Delta}_{{S_{1}}}}\right\rVert_{F}
\end{align}

\begin{align}
    & \lambda_{{L_2}}{C_2}({\hat{\Delta}_{{L_{2}}}}^{A_2}, {\hat{\Delta}_{{S_{2}}}}^\mathbb{N}) \leq  \sqrt{2 R_2} \lambda_{{L_2}}  \left \lVert {\hat{\Delta}_{{L_{2}}}} \right\rVert_{F} + \sqrt{s_2} \lambda_{{S_2}} \left \lVert {\hat{\Delta}_{{S_{2}}}}\right\rVert_{F}
\end{align}

Combining these two equations above with \ref{eq22}, and ignoring the unnecessary constants, we get the following:

\begin{align}
    &\Bigg[\left \lVert {\hat{\Delta}_{{L_{1}}}}\right\rVert _{F}^{2} +  \left \lVert {\hat{\Delta}_{{S_{1}}}}\right\rVert _{F}^{2} +  \left \lVert {\hat{\Delta}_{{L_{2}}}}\right\rVert _{F}^{2} +  \left \lVert {\hat{\Delta}_{{S_{2}}}}\right\rVert _{F}^{2}\Bigg] 
    \preceq
    \sqrt{ R_1} \lambda_{{L_1}}  \left \lVert {\hat{\Delta}_{{L_{1}}}} \right\rVert_{F} + \sqrt{s_1} \lambda_{{S_1}} \left \lVert {\hat{\Delta}_{{S_{1}}}}\right\rVert_{F}  \nonumber \\ 
    & \hspace{8.4cm} + \sqrt{ R_2} \lambda_{{L_2}}  \left \lVert {\hat{\Delta}_{{L_{2}}}} \right\rVert_{F} + \sqrt{s_2} \lambda_{{S_2}} \left \lVert {\hat{\Delta}_{{S_{2}}}}\right\rVert_{F}
\end{align}

From the above equation, we can write the following,

\begin{align}
    & \left \lVert {\hat{\Delta}_{{L_{1}}}}\right\rVert _{F}^{2} +  \left \lVert {\hat{\Delta}_{{S_{1}}}}\right\rVert _{F}^{2} +  \left \lVert {\hat{\Delta}_{{L_{2}}}}\right\rVert _{F}^{2} +  \left \lVert {\hat{\Delta}_{{S_{2}}}}\right\rVert _{F}^{2}
    \preceq
     R_1 \lambda_{{L_1}}^2 +  s_1 \lambda_{{S_1}}^2 + R_2 \lambda_{{L_2}}^2 +  s_2 \lambda_{{S_2}}^2
\end{align}

This completes the proof of the lemma.

\subsection*{\textit{Proof of Theorem \ref{thm:34}}}

At first, we establish that $\lambda_{{S_1}} \geq  4 \left \lVert {\mathcal{D}_1} \right\rVert_{\infty} + \frac{4\gamma \alpha_1}{\sqrt{d_1 d_1}}$  and $\lambda_{{S_2}} \geq  4 \left \lVert {\mathcal{D}_2} \right\rVert_{\infty} + \frac{4\gamma \alpha_2}{\sqrt{d_2 d_2}}$ are satisfied with high probability. \\

Recalling the notation from Section \ref{theo}, and applying Proposition 2.4(b) in \cite{basu2015regularized} to the matrices $\mathbb{E}_1$ and $Y_{-1}^{(1)}$, we can say that that there exists a constant $c >0$ such that for any $u, v \in \mathbb{R}^{d_1}$ with $ \left \lVert u \right \rVert \leq 1$,$ \left \lVert v \right \rVert \leq 1$ and for any $\eta >0$, we get
\begin{align}
    & P \Bigg[ \left \lvert u^T \Big( \frac{{E_1} {Y_{-1}^{(1)^{T}}}}{T}  \Big) v \right \rvert > 2\pi Q_1 \eta \Bigg]
    \leq
    6 \hspace{0.1cm} \text{exp}[-cT min \{\eta, \eta^2\}]
\end{align}

Taking the same approach as in the proof of Proposition 4.3 in \cite{basu2015regularized}, we take the union bound over the $d_1^2$ possible choices of $u \in \{e_1, e_2, \dots e_{d_{1}} \}$ and $v \in \{e_1, e_2, \dots e_{d_{1}} \}$ to get the following:

\begin{align}
    & P \Bigg[ \frac{\left \lVert {E_1}{Y_{-1}^{(1)^{T}}}  \right \rVert _{\infty}}{T} > 2\pi Q_1 \eta \Bigg]
    \leq
    6 \hspace{0.1cm} \text{exp}[-cT min \{\eta, \eta^2\} + 2 \text{log} (d_1)]
\end{align}

Now, we take $\eta = \sqrt{\frac{2\text{log}(d_1)}{T}}$, to get,

\begin{align}
    &  P \Bigg[ \frac{\left \lVert {E_1}{Y_{-1}^{(1)^{T}}}  \right \rVert _{\infty}}{T}> 2\pi Q_1 \sqrt{\frac{2\text{log}(d_1)}{T}} \Bigg]
    \leq
    6 \hspace{0.1cm} \text{exp}[-c_1 \text{log}(d_1)]
\end{align}
for a suitably chosen constant $c_1$. Thus, we choose $\lambda_{{S_{1}}} = k_1 Q_1 \sqrt{\frac{2\text{log}(d_1)}{T}} + \frac{4 \gamma \alpha_{1}}{\sqrt{d_{1}d_{1}}}$, for some suitably chosen constant $k_1$. Following a similar reasoning, it can be shown that 
\begin{align}
    &  P \Bigg[ \frac{\left \lVert {E_2}^T{Y_{-1}^{(2)}}  \right \rVert _{\infty}}{T}> 2\pi Q_2 \sqrt{\frac{2\text{log}(d_2)}{T}} \Bigg]
    \leq
    6 \hspace{0.1cm} \text{exp}[-c_2 \text{log}(d_2)]
\end{align}
for a suitably chosen constant $c_2$. So we choose $\lambda_{{S_{2}}}$ as $k_2 Q_2 \sqrt{\frac{2\text{log}(d_2)}{T}} + \frac{4 \gamma \alpha_{2}}{\sqrt{d_{2}d_{2}}} $ for some suitable chosen constant $k_2$.

\vspace{0.1in}
Now we establish that, $\lambda_{L_1} \geq 4 \left \lVert \mathcal{D}_1 \right\rVert_{sp}$ and $\lambda_{L_2} \geq 4 \left \lVert \mathcal{D}_2 \right\rVert_{sp}$ are satisfied with high probability. To that end, let $\mathcal{S}^{d_1-1}$ denote the unit ball for $R^{d_1}$. We discretize this unit ball using $\epsilon$-net $\mathcal{N}$ with cardinality at most $(1+\frac{2}{\epsilon})^{d_1}$. Now following the same argument as in Lemma F.2 of \cite{basu2015regularized}, for small enough $\epsilon > 0$,
\begin{equation}
    \underset{u \in \mathcal{S}^{d_1-1}, v \in \mathcal{S}^{d_1-1}}{sup} \lvert u^\prime \frac{(\mathbb{E}_1Y_{-1}^{(1)^T})}{T}v\lvert \leq k \underset{u \in \mathcal{N}, v \in \mathcal{N}}{sup} \lvert u^\prime \frac{(\mathbb{E}_1Y_{-1}^{(1)^T})}{T}v\lvert  
\end{equation}
for some suitable chosen constant $k$. Now, as before, taking union bound over $(1+\frac{2}{\epsilon})^{2d_1}$ choices of $u$ and $v$ we get,
\begin{equation}
    Pr\{\frac{\norm{\mathbb{E}_1Y_{-1}^{(1)^T}}_{sp}}{T} > 2\pi \text{ } k\eta Q_1\} \leq 6\text{ }exp[-cT \text{ }min\{\eta^2,\eta\}+ 2d_1\log(1+\frac{2}{\epsilon})] 
\end{equation}
Hence we choose $\eta=\sqrt{\frac{c_12d_1\log(1+\frac{2}{\epsilon})}{cT}}$ and the above equation boils down to 
\begin{equation}
    Pr\{\frac{\norm{\mathbb{E}_1Y_{-1}^{(1)^T}}_{sp}}{T} > 2\pi \text{ } k\sqrt{\frac{c_12d_1\log(1+\frac{2}{\epsilon})}{cT}} Q_1\} \leq 6\text{ }exp[-c_3d_1] 
\end{equation}

for a suitable chosen constant $c_3$. So we choose $\lambda_{L_1}=k_1^*Q_1\sqrt{\frac{2d_1}{T}}$, for a suitable chosen constant $k_1^*$. Following a similar reasoning, it can be shown that 
\begin{equation}
    Pr\{\frac{\norm{\mathbb{E}_2^TY_{-1}^{(2)}}_{sp}}{T} > 2\pi \text{ } k\sqrt{\frac{c_12d_2\log(1+\frac{2}{\epsilon})}{cT}} Q_2\} \leq 6\text{ }exp[-c_4d_2] 
\end{equation}

for a suitable chosen constant $c_4$. So we choose $\lambda_{L_2}=k_2^*Q_2\sqrt{\frac{2d_2}{T}}$, for a suitable chosen constant $k_2^*$. Now the proof of the theorem follows by using these choices of the regularizer parameters and putting the same in the bound obtained in Lemma \ref{lemma:33}.

\end{appendices}

\end{document}